\documentclass[journal, twocolumn]{IEEEtran}
\usepackage{cite}
\usepackage{amsmath,amsfonts}
\usepackage{algorithmic}
\usepackage{algorithm}
\usepackage{array}
\usepackage[caption=false,font=normalsize,labelfont=sf,textfont=sf]{subfig}
\usepackage{textcomp}
\usepackage{stfloats}
\usepackage{url}
\usepackage{verbatim}
\usepackage{graphicx}
\usepackage{relsize}
\usepackage[utf8]{inputenc}
\usepackage{mathptmx}
\usepackage{enumitem}
\usepackage{hyperref}
\usepackage{adjustbox}
\usepackage{multirow}
\usepackage{xcolor}
\usepackage{caption}

\newlength{\barW}
\newlength{\barH}
\newcommand{\fillbar}[1]{%
  \begin{picture}(10,7)(0,-1)
    \put(0,0){\line(1,0){10}}   % bottom
    \put(0,5){\line(1,0){10}}  % top
    \put(0,0){\line(0,1){5}}    % left
    \put(10,0){\line(0,1){5}}  % right
    \put(0,0){\color{black}\rule{#1\barW}{\barH}}%
  \end{picture}%
}

\newcommand{\emptycircle}{\fillbar{0}}
\newcommand{\thirdpie}{\fillbar{0.33}}
\newcommand{\twothirdspie}{\fillbar{0.66}}
\newcommand{\fullcircle}{\fillbar{1}}

\def\BibTeX{{\rm B\kern-.05em{\sc i\kern-.025em b}\kern-.08em
    T\kern-.1667em\lower.7ex\hbox{E}\kern-.125emX}}
    
\begin{document}
%\history{Date of publication xxxx 00, 0000, date of current version xxxx 00, 0000.}
%\doi{10.1109/ACCESS.2017.DOI}

\title{How Feasible are Passive Network Attacks on 5G Networks and Beyond?\\A Survey}
\author{\uppercase{Atmane Ayoub Mansour Bahar},~\IEEEmembership{Graduate Student Member,~IEEE},
\uppercase{Andr\'es Alay\'on Glazunov}, and
\uppercase{Romaric Duvignau},~\IEEEmembership{Senior Member,~IEEE}%
\thanks{Atmane Ayoub Mansour Bahar and Romaric Duvignau are with the Department of Computer Science and Engineering, Chalmers University of Technology and the University of Gothenburg, SE-412 96 Gothenburg, Sweden (e-mail: atmane@chalmers.se; duvignau@chalmers.se).}%
\thanks{Andr\'es Alay\'on Glazunov is with the Department of Science and Technology, Link\"oping University, Norrk\"oping Campus, 601 74 Norrk\"oping, Sweden (e-mail: andres.alayon.glazunov@liu.se).}%
\thanks{Corresponding author: Atmane Ayoub Mansour Bahar (e-mail: atmane@chalmers.se).}%
}

\markboth
{Mansour Bahar \textit{et al.}}
{How Feasible Are Passive Network Attacks on 5G Networks and Beyond? A Survey}

\maketitle

\begin{abstract}
Privacy concerns around 5G, the latest generation of mobile networks, are growing, with fears that its deployment may increase exposure to privacy risks. This perception is largely driven by the use of denser deployments of small antenna systems, which enable highly accurate data collection at higher speeds and closer proximity to mobile users. At the same time, 5G’s unique radio communication features can make the reproduction of known network attacks more challenging. In particular, passive network attacks, which do not involve direct interaction with the target network and are therefore nearly impossible to detect, remain a pressing concern. Such attacks can reveal sensitive information about users, their devices, and active applications, which may then be exploited through known vulnerabilities or spear-phishing schemes. This survey examines the feasibility of passive network attacks in 5G and beyond (B5G/6G) networks, with emphasis on two major categories: information extraction (system identification, website and application fingerprinting) and geolocation (user identification and position tracking). These attacks are well documented and reproducible in existing wireless and mobile systems, including short-range networks (IEEE 802.11) and, to a lesser extent, LTE. Current evidence suggests that while such attacks remain theoretically possible in 5G, their practical execution is significantly constrained by beamformed control channels, PDCP ciphering, and the specialized capture equipment required. Available evidence does not establish general practical infeasibility of passive attacks in B5G/6G.


\end{abstract}

%\titlepgskip=-15pt

%\maketitle

\section{Introduction}
\label{sec:intro}
\begin{figure}[t]
    \centering
    \includegraphics[width=\linewidth]{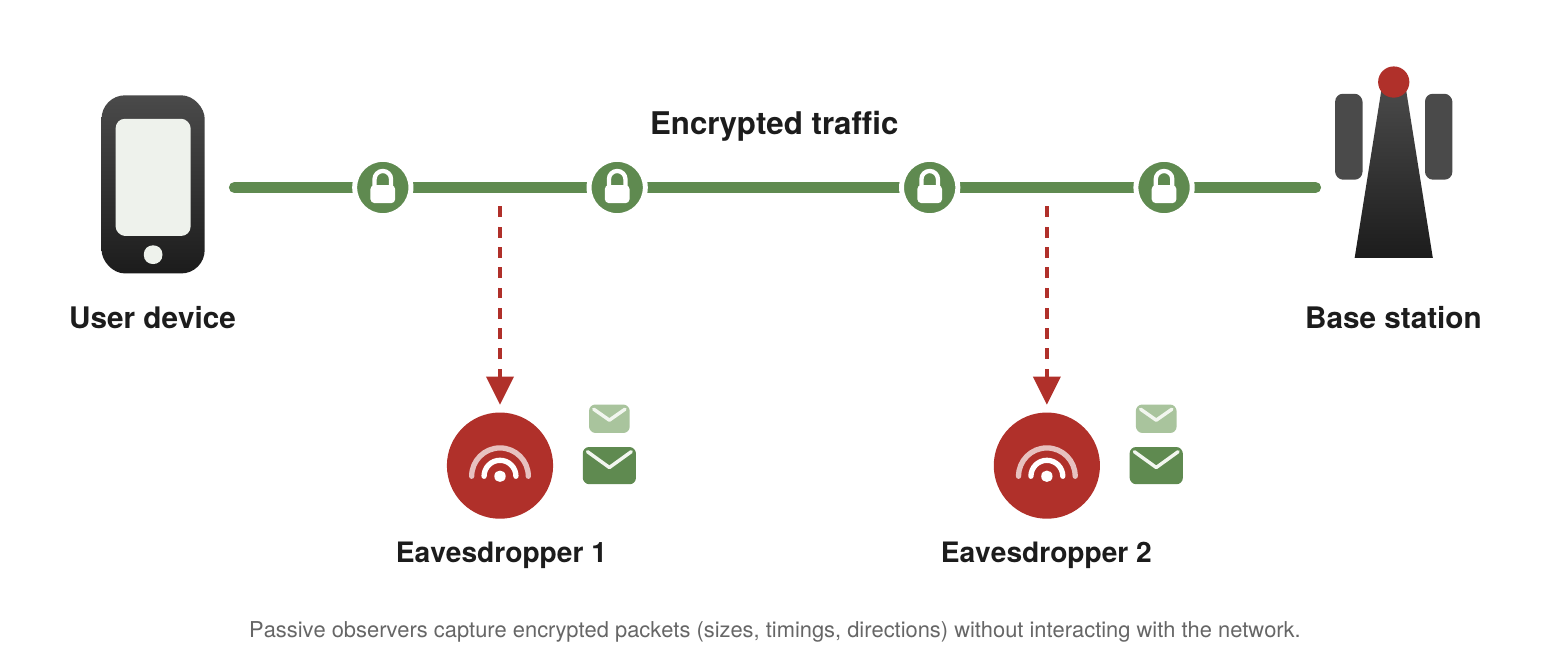}
    \caption{Eavesdropping on cellular communications at the edge.}
    \label{fig:5g}
\end{figure}

Privacy protection~\cite{senivcar2003privacy} has become an increasingly important concern in the past decade, reinforced by the advent of the General Data Protection Regulation (GDPR) and similar legislation. In particular, there is growing concern that 5G~\cite{kaska2019huawei,ahmad2019security}, the latest generation of mobile networks, is more vulnerable to privacy risks and eavesdropping (see Fig.~\ref{fig:5g} for an illustration). This belief is largely driven by the denser deployment of small antenna systems positioned closer to mobile users, which enable faster, more accurate data transmission with lower latency~\cite{gadgethacksMajorPrivacy, rabine2022rise, tomasin2021location, acsGivesGovernment}. At the same time, through beamforming and massive MIMO (Multiple Input Multiple Output), 5G’s New Radio communications are much more directional than in previous generations, which makes the reproduction of known attacks significantly more challenging in this environment~\cite{mezzavilla2018end}. Consequently, it remains unclear whether 5G and beyond 5G (B5G/6G) networks pose greater or lesser privacy risks compared with other forms of wireless communication and earlier cellular standards.

In parallel, the volume of data generated by internet-connected devices has sharply increased. Combined with greater data availability and advances in machine learning (ML) technologies, this growth enables the identification of individual mobile users and their behaviors with high accuracy~\cite{reed2016leaky,reed2017identifying,conti2018dark,acar2020peek,meneghello2020smartphone}. Such information mining increasingly relies on subtle user data, particularly encrypted network traces. These traces, which act as a side-channel of wireless communication, are continuously produced, difficult to control, and therefore an attractive target. Analyzing them, a process referred to as \textit{network trace analysis}, enables the design of entirely \textit{passive attack schemes}, which do not require any interaction with the attacked network. Because these attacks only extract information from network logs, they are practically impossible for victims to detect. Nevertheless, they can create serious security risks, as the extracted information may be exploited to target known or even zero-day vulnerabilities in software or network protocols. Even more concerning is that such passive attacks are relatively easy to set up, potentially enabling individuals without advanced networking expertise to acquire private information about victims. This opens the door to spear-phishing schemes, such as sending fake job offers after detecting traffic from a job-seeking application, proposing fraudulent medical treatments, or even attempting direct blackmail based on visited websites~\cite{miller2014know}.

\subsection{Challenges for attackers}
\begin{figure}[t]
    \centering
    \includegraphics[width=\linewidth]{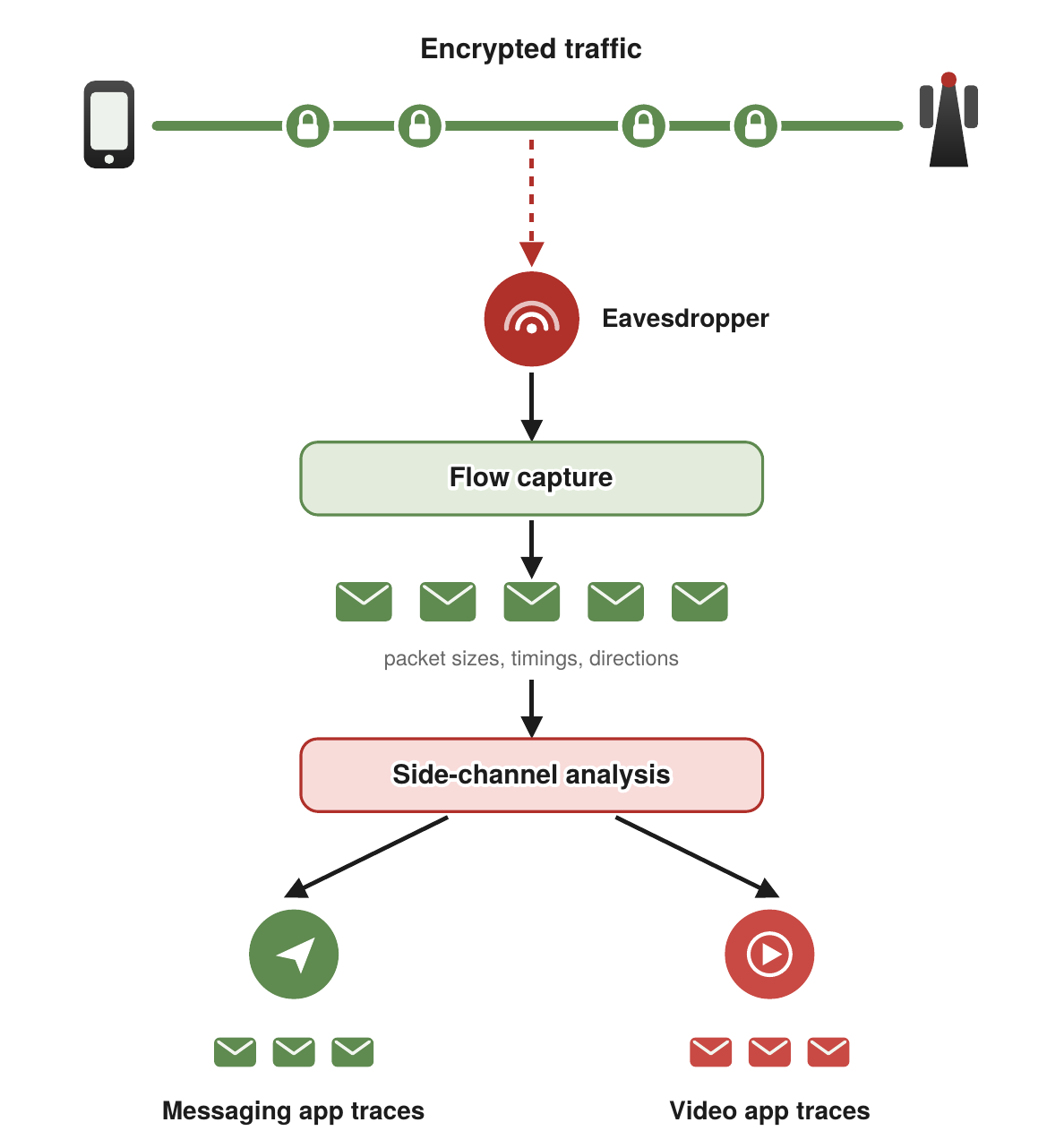}
    \caption{Illustration of encrypted traffic classification from captured traces.}
    \label{fig:tc}
\end{figure}

The main objectives of packet trace analysis include traffic classification (see Fig.~\ref{fig:tc} for an illustration), user identification and localization, website fingerprinting, application recognition, and data-content inference. In general, only acquiring information from side-channel data is inherently difficult. When communication is encrypted at the link layer, information extraction becomes even more challenging. Eavesdropping on cellular communications is particularly demanding not only because of such encryption, but also due to the limited availability of commercially sold capture interfaces, the scarcity of open-source tools, the noisier characteristics of wireless channels, and the higher variability of network traces.  

\subsection{Motivations for this survey} 
Packet trace analysis over encrypted communication channels poses serious societal challenges. 

This work surveys the current landscape of existing and potential Passive Network Attacks (PNAs), with the aim of informing the design of countermeasures before such attacks are widely deployed. A careful study of these possible attacks can also encourage the early adoption of necessary updates in wireless protocols and standards. The primary motivation of this survey is to provide a clearer understanding of privacy concerns in 5G and B5G networks~\cite{khan2018defeating,norrman2016protecting}, thereby addressing public apprehension surrounding their deployment. By offering a qualitative analysis of privacy risks, this work seeks to reduce popular concerns about new wireless technologies, which in turn can help increase public acceptance and accelerate the development and deployment of both current and future network generations.  

\subsection{Contributions} 
By surveying 44 works from literature, this work investigates the feasibility of PNAs on 5G and B5G networks. Particular emphasis is placed on information extraction related to users' devices (e.g., system identification, website fingerprinting, and application fingerprinting) and on users' geographical position~\cite{yang2016passive} (e.g., device tracking). The survey primarily considers attacks that have already been demonstrated in wireless networks, either in short-range technologies or LTE cellular systems. However, it remains unclear whether these attacks can be easily adapted to 5G, since the new radio characteristics of 5G integrate beamforming into initial access and control-channel transmission, which can make an eavesdropper's success depend more strongly on its position relative to the target and the gNodeB~\cite{lautenbach2019preliminary}. It is also uncertain whether they will transfer to B5G, which builds upon the capabilities of today’s 5G environments~\cite{gsmaintelligenceTechnologyWhite}. 

Unlike prior surveys, which either (i) classify side-channel/traffic-analysis techniques independent of the underlying access technology~\cite{conti2018dark, rahbari2015secrecy, spreitzer2017systematic, salman2020review, salman2021data, kumar2021smartphone, papadogiannaki2021survey}, or (ii) survey 5G/B5G security broadly without assessing specific attack reproducibility~\cite{cao2019survey, khan2019survey, Sandeepa_2022, ramezanpour2022securityprivacyvulnerabilities5g6g}, or (iii) discuss eavesdropping or physical-layer threats in 5G/6G at a conceptual level without a systematic feasibility assessment~\cite{s24175523, wani2024security, saeed2025comprehensive, DEVI2025100891} (see Table \ref{tab:survey_comparison}), this survey is, to the best of our knowledge, the first to combine a fixed taxonomy of demonstrated passive attacks with an explicit, quantifiable feasibility classification (Table \ref{tab:pna-feasibility}) tied to the specific radio and protocol mechanisms (beamforming gain, RNTI search-space size, PDCP encryption scope) that determine whether each attack transfers to 5G and B5G.

\subsection{Plan}
Section \ref{sec:pre} introduces the key concepts used throughout this work: Section \ref{subsec:attack-types} defines the taxonomy of attack types (PNAs, side-channels, traffic analysis, eavesdropping, WF, VF), Section \ref{subsec:pre-tools} summarizes the tools and techniques used by network administrators to detect or analyze such attacks, and Section \ref{subsec:5g_com} details the characteristics of 5G and B5G/6G communications most relevant to PNA feasibility, including the specific challenges of eavesdropping on 5G traffic. Section \ref{sec:related-work} positions this survey relative to prior work: Section \ref{subsec:surveys} compares it against related surveys on traffic analysis and 5G/B5G security (Table \ref{tab:survey_comparison}), while Section \ref{subsec:passive_attacks} reviews the broader literature on PNAs across Wi-Fi, LTE, and early 5G/6G systems. Section \ref{sec:methodology} describes our methodology: the criteria used to select the 44 works surveyed (Section \ref{subsec:data-selection}), a detailed review of each (Section \ref{subsec:data}, summarized in Table \ref{tab:pna-summary}), and a classification of these works by input data type, i.e., packet sequences, bidirectional flows, or control-plane/physical-layer signals (Section \ref{subsec:data-classification}). Section \ref{sec:feasbility} is our core contribution: it introduces the physical-, protocol-, and resource-layer factors that govern PNA feasibility in 5G (Section \ref{subsec:factors}), classifies each of the 44 surveyed works into one of four feasibility classes for 5G reproduction (Section \ref{subsec:dempónstrability}, Table \ref{tab:pna-feasibility}), discusses the resulting trends and their implications for B5G/6G (Section \ref{subsec:discussion}, Table \ref{tab:feas}), and examines emerging attack surfaces, such as Ambient IoT, network power-saving mechanisms, physical-layer security, and positioning, that fall outside our core scope but are worth flagging explicitly (Section \ref{subsec:aiot}). Finally, Section \ref{sec:conclusion} concludes the paper, states its limitations, and outlines directions for future work.

\section{Preliminaries} \label{sec:pre}
We present in this section definitions of important concepts tackled in this work.

\subsection{Types of Attacks} \label{subsec:attack-types}
We start by explaining some types of attacks that are repeatedly mentioned throughout this paper. This is not an exhaustive taxonomy of network attacks; rather, we define here the specific subset of attack categories that recur in the 44 works surveyed in Section \ref{sec:methodology}, following definitions adapted from prior side-channel and traffic-analysis taxonomies~\cite{rahbari2015secrecy, spreitzer2017systematic, papadogiannaki2021survey}.

\paragraph{Passive Network Attacks}   
Attacks that monitor or intercept network traffic without altering it or interacting with the network in any detectable way, aiming to gather information covertly~\cite{rahbari2015secrecy, spreitzer2017systematic}. Their counterpart, "active network attacks," involve interactions with the network (e.g., injected packets, rogue base stations soliciting connections) which can in principle be detected via, e.g., Intrusion Detection Systems.

\paragraph{Eavesdropping} The unauthorized interception of private communications more broadly, such as listening to voice calls or reading transmitted data~\cite{papadogiannaki2021survey}. Eavesdropping is a purely observational goal, and it can be carried out either passively (as in a PNA, with no interaction with the target network) or, in some cellular contexts, semi-actively; for instance, via a rogue base station that solicits a device to connect before intercepting its traffic. \emph{Every PNA is therefore a form of eavesdropping, but not every eavesdropping attack qualifies as a PNA}; this survey restricts its scope to the passive subclass, since the strictly non-interactive nature of PNAs is what makes them practically undetectable by the victim or network operator (see Section \ref{subsec:data-selection} for the corresponding threat model).
We nonetheless include two relay-based works \cite{cheng2023watching, wang2026app} as reference points, since they bound what an adversary gains once it can interpose itself on the link; they are marked FLW* in Table \ref{tab:pna-feasibility} and are not strictly passive.

\paragraph{Communication side-channels} Exploitation of indirect information to infer sensitive data without direct access to content~\cite{spreitzer2017systematic}. When the main channel is encrypted, packet traces form an exploitable side-channel (e.g. packet sizes and timestamps). Other possible side-channels on mobile devices could be energy consumption and voltage.
    
\paragraph{Traffic Analysis} The observation and examination of network traffic patterns to deduce communication behaviors, relationships, or identities~\cite{conti2018dark, papadogiannaki2021survey}.

%% Might group eavesdropping within communication side-channels
\paragraph{Website Fingerprinting (WF)} Inferring which website a user is visiting by analyzing encrypted traffic patterns~\cite{juarez2014critical}. It usually requires the attacker to build a database of, e.g., the 200 most popular websites to identify captured traffic~\cite{rimmer2018automated}.

\paragraph{Video Fingerprinting (VF)} Identifying specific video content being streamed by examining traffic features like bitrate or packet timing~\cite{reed2016leaky, reed2017identifying}. Just like WFs, a database of fingerprints is usually required.

\subsection{Tools and Techniques of Network Administrators} \label{subsec:pre-tools}
We move next to different techniques and tools used by network administrators to detect attacks or analyse attack traces.

\paragraph{Traffic Classification (TC)} Categorizing network flows\footnote{We define a ``flow" as a structured aggregation of packets that belong to the same communication session or connection.} (e.g., VoIP, web, streaming) to enforce policies, optimize routing, or detect anomalies using observed traffic characteristics, which varies depending on whether the traffic is encrypted or not.

\paragraph{Deep Packet Inspection (DPI)} Traditional method to inspect payload data for security, policy enforcement, or traffic shaping.

\paragraph{Quality of Service (QoS)} Ensuring reliable network performance by prioritizing critical traffic and managing bandwidth allocation.

\subsection{Characteristics of (B)5G communications} \label{subsec:5g_com}
We present in this section the main characteristics of 5G and B5G networks and explore their implications for the feasibility of PNAs.

\subsubsection{5G New Radio standard}
The 5G wireless technology is the basis of a new kind of wireless network that enables an omnipresent connectivity to virtually link everyone and everything including machines, objects, and devices. 5G wireless networks are built to deliver higher multi-Gbps peak data streams, ultra low latency, more reliability, vast network capacity, and improved availability, uniformly to more users~\cite{dahlman20205g}. 

5G New Radio (5G NR) is a new air interface deployed for 5G. NR provides forward compatibility within NR and interworks with LTE via NSA (EN-DC\footnote{EN-DC (E-UTRAN New Radio – Dual Connectivity.) is 5G NSA feature that allows a device to connect simultaneously to a 4G LTE eNodeB (anchor for control and signaling) and a 5G NR gNodeB (for additional high-speed data), combining both for higher throughput.}), however, it is not air-interface backward-compatible with LTE. 

\begin{figure*}[t]
    \centering
    \includegraphics[width=1\linewidth]{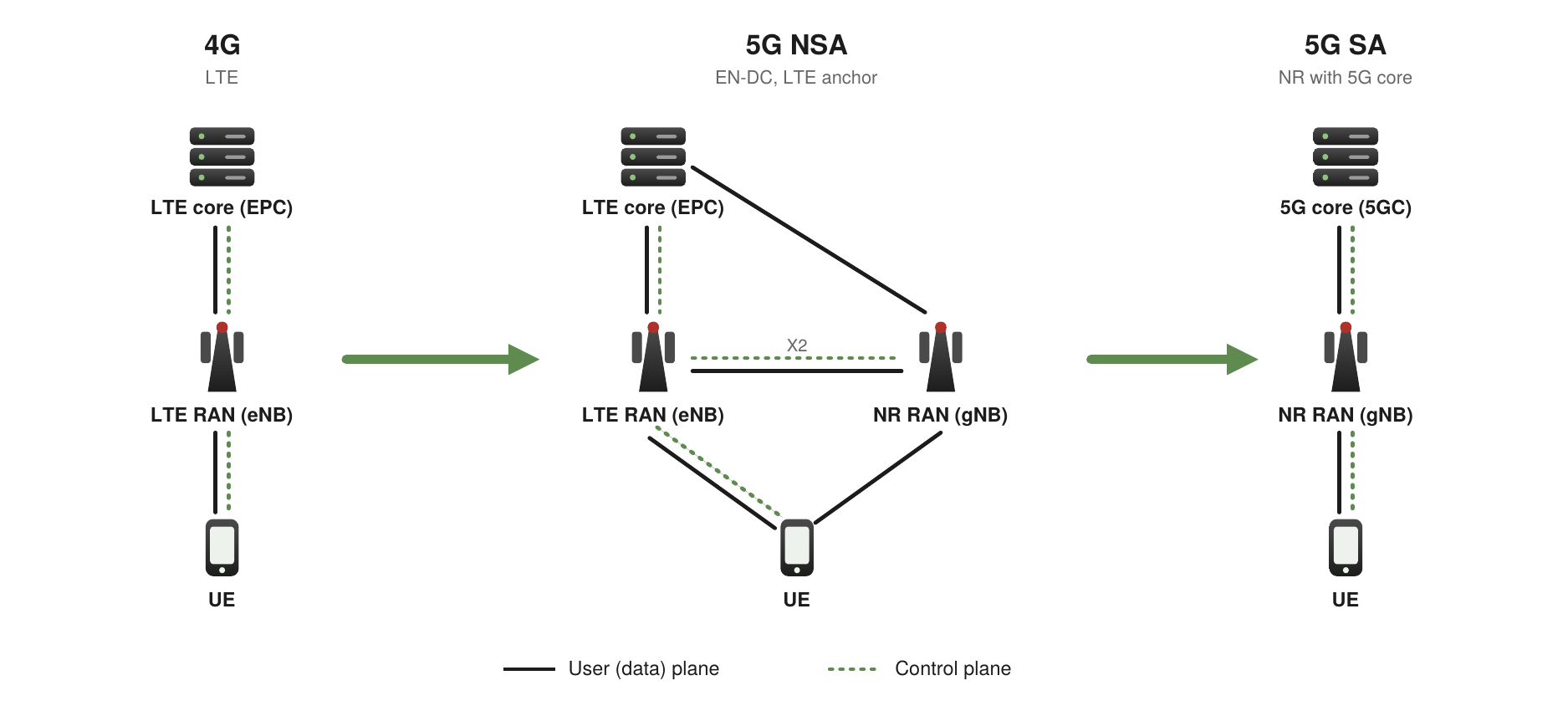}
    \caption{Evolution of the network architecture from 4G to 5G.}
    \label{fig:5gnr}
\end{figure*}

5G NR has two broad deployment modes: NSA (EN-DC), where LTE anchors the control plane and NR primarily adds user plane capacity, and \textit{Standalone} (SA), where NR carries both control and user planes on a 5G core. Operators may use Dynamic Spectrum Sharing (DSS) to run LTE and NR in the same band though this mode is optional and deployment-specific.
As technology matures, the SA mode operates with the 5G core network and hence at both the control and user planes. In the SA mode, the 5G Packet Core architecture is fully used instead of the 4G Evolved Packet Core run in the 4G LTE network (see Fig. \ref{fig:5gnr} for a summary.)

The foundational elements of 5G NR are the following. NR reuses Orthogonal Frequency Division Multiplexing (OFDM) and Cyclic Prefix OFDM (CP-OFDM) waveforms (also used in LTE and Wi-Fi), but it does not reuse Wi-Fi protocols. %Existing 4G LTE and WiFi technology is reused. 
Namely, wave forms based on OFDM and multiple access techniques are optimized. A common flexible framework will enable efficient multiplexing of diverse 5G services and provide forward compatibility for future services. Advanced wireless technologies deliver new levels of performance and efficiency that enables the wide range of 5G services. 5G services are categorized into three pillars, namely enhanced Mobile Broadband (eMBB) delivering high throughput; ultra-reliable and Low-latency Communications (URLLC) for high reliability and availability (not security); and massive Machine Type Communications (mMTC) for low-cost, low-energy devices with small data volumes at mass scale. 

While the main enablers of 5G include larger bandwidths in the millimeter wave (mmWave) frequency bands, densification via small-cell HetNets, and massive MIMO with narrow beams~\cite{khwandah2021massive}, in practice 5G capacity is delivered by a layered spectrum mix across low-, mid-, and mmWave bands, with mid-band carrying most of today’s load and mmWave supplying extreme peak rates where deployed.
In addition, due to the shortcomings Orthogonal Multiple Access (OMA) lacking sufficient spectral efficiency to handle the foreseen unprecedented increase of data traffic Non-Orthogonal Multiple Access (NOMA) has become a promising remedy. (Note, however, that NR does not standardize general power-domain NOMA in current releases). Furthermore, NR relies on Frequency Division Duplex (simultaneous Uplink/Downlink on separate bands) and Time Division Duplex (time-separated Uplink/Downlink). In-band full-duplex on the same frequency is being studied but is not required or widely deployed, however, some researchers advocate it for potential throughput gains.

\subsubsection{Challenges of eavesdropping 5G communications} \label{subsec:challenges5g}
Eavesdropping on 5G communications is fundamentally more constrained than on earlier wireless systems, not merely because commodity capture hardware is unavailable, but because 5G NR was architecturally designed around directional transmission, dynamic spectrum access, and mandatory link-layer confidentiality. The 3GPP NR physical and MAC layer specification~\cite{3gpp38300} defines beam-based access procedures (beam sweeping, beam management, and beam-specific control resource sets) as core parts of initial access and scheduling, rather than optional enhancements, and the 5G security architecture~\cite{3gpp33501} requires support for PDCP-layer ciphering, which is activated according to the operator's security policy, and mandates ciphering of most control-plane signaling. Together, these design choices mean that the challenge for a passive eavesdropper is one of degree and mechanism, not merely of device availability: even an attacker with capture-capable hardware faces a system engineered to minimize unintended radiation and to conceal scheduling and content information from anyone other than the intended receiver. For illustrative contrast, Wi-Fi's simpler, largely omnidirectional PHY/MAC and its long history of commodity monitor-mode NICs make it a useful baseline for comparison, but the underlying difficulty of 5G eavesdropping stems from these cellular-specific constraints (beamforming, NR numerology and dynamic spectrum sharing, and PDCP encryption) rather than from the mere absence of an equivalent "monitor mode".\footnote{Note that mmWave initial access is also vulnerable to targeted jamming~\cite{mezzavilla2018end} during initial access and beam tracking; this vulnerability is largely mmWave-centric and scenario-dependent (sub-6 GHz NR behaves differently). This is an active-attack concern outside this survey's passive-attack scope and is mentioned only to distinguish it from the eavesdropping risks discussed above.}

Capturing and demodulating 5G signals passively requires a Software Defined Radio (SDR, a radio device that can switch between different wireless communication standards using software, without needing new hardware~\cite{akeela2018software}) front-end whose instantaneous bandwidth matches the observation task. Capturing a full NR carrier requires covering its entire channel bandwidth, up to 100 MHz in FR1 and 400 MHz in FR2. Many control-plane attacks, however, need only narrower portions:
\begin{itemize}
    \item the SS/PBCH block (SSB) occupies 20 resource blocks (7.2 MHz at 30 kHz subcarrier spacing);
    \item CORESET 0 spans 24 to 96 resource blocks (8.64 to 34.56 MHz at 30 kHz);
    \item a UE's active bandwidth part may be configured narrower than the carrier.
\end{itemize}
Real-time demodulation must also keep pace with sub-millisecond slot timing, and some multi-antenna strategies need several synchronized RF chains. Entry-level SDRs (e.g., RTL-SDR, HackRF) offer at most about 20 MHz of instantaneous bandwidth. Mid-range units such as the USRP B210 have 56 MHz of analog bandwidth, and standard UHD operation limits their master clock to 30.72 MHz in dual-channel mode \cite{ettusb2x0}. Such units can observe the SSB, CORESET 0, or a narrow bandwidth part, but not a full 100 MHz carrier, which requires higher-end platforms such as USRP X310/N310-class hardware (see Table \ref{tab:tools}). 
While open-source cellular stacks such as srsRAN~\cite{srsranSrsRANProject} and OpenAirInterface~\cite{oai} provide a foundation, they are designed as protocol endpoints (UE or gNodeB) rather than promiscuous sniffers, and adapting them for passive capture requires substantial custom engineering. Recent academic efforts illustrate both the feasibility and the difficulty of closing this gap: Wan et al.~\cite{wan2024nr} built a dedicated 5G SA telemetry decoder (NR-Scope) to extract PDCCH/DCI scheduling information from a live network, and Ludant et al.~\cite{ludant20235g} developed 5GSniffer, to our knowledge the first open-source passive 5G control-channel sniffer, to track UE presence via decoded control messages. Both required extensive manual calibration and purpose-built decoding pipelines, underscoring that no turnkey, general-purpose 5G passive monitoring stack yet exists, despite active efforts in the research community.

Beamforming compounds this hardware barrier at the signal level. Although 5G beamforming is intended to improve spectral efficiency and throughput rather than to provide confidentiality, it concentrates radiated energy toward the scheduled UE. Whether an eavesdropper can decode a given transmission then depends on the transmit array pattern, its own position and receive gain, and the propagation environment. It may obtain sufficient SINR through the main lobe if it lies within the beam's finite width, through a side lobe if it is close enough or uses a high-gain receiver, or through multipath reflections. Reliable capture is therefore geometry- and configuration-dependent rather than guaranteed or excluded; we formalize this in Section \ref{subsec:phy-constraints}.

Even when a usable signal is captured, PDCP-layer ciphering~\cite{3gpp33501} prevents an external eavesdropper from separating individual user flows without the decryption keys, so any traffic recovered from an over-the-air capture necessarily mixes packets from concurrent, unrelated applications and users sharing the same radio resources. This has two compounding consequences for the classification methods used in this survey. First, models must tolerate substantially noisier, unlabeled input than the clean, single-flow traces typically used in Wi-Fi- or LTE-based studies. Second, because packet-timing and size characteristics vary by device and chipset, models trained on one device or vendor generalize poorly to others, a limitation independently reported in encrypted-traffic classification studies both in fixed networks~\cite{aceto2019mobile, aceto2020toward} and under noisy or heterogeneous conditions~\cite{ismailaj2021deep}; scaling any 5G passive-attack pipeline across device populations therefore remains an open, resource-intensive problem rather than a solved one.

\subsubsection{Beyond 5G characteristics}
\begin{figure}[t]
    \centering
    \includegraphics[width=1\linewidth]{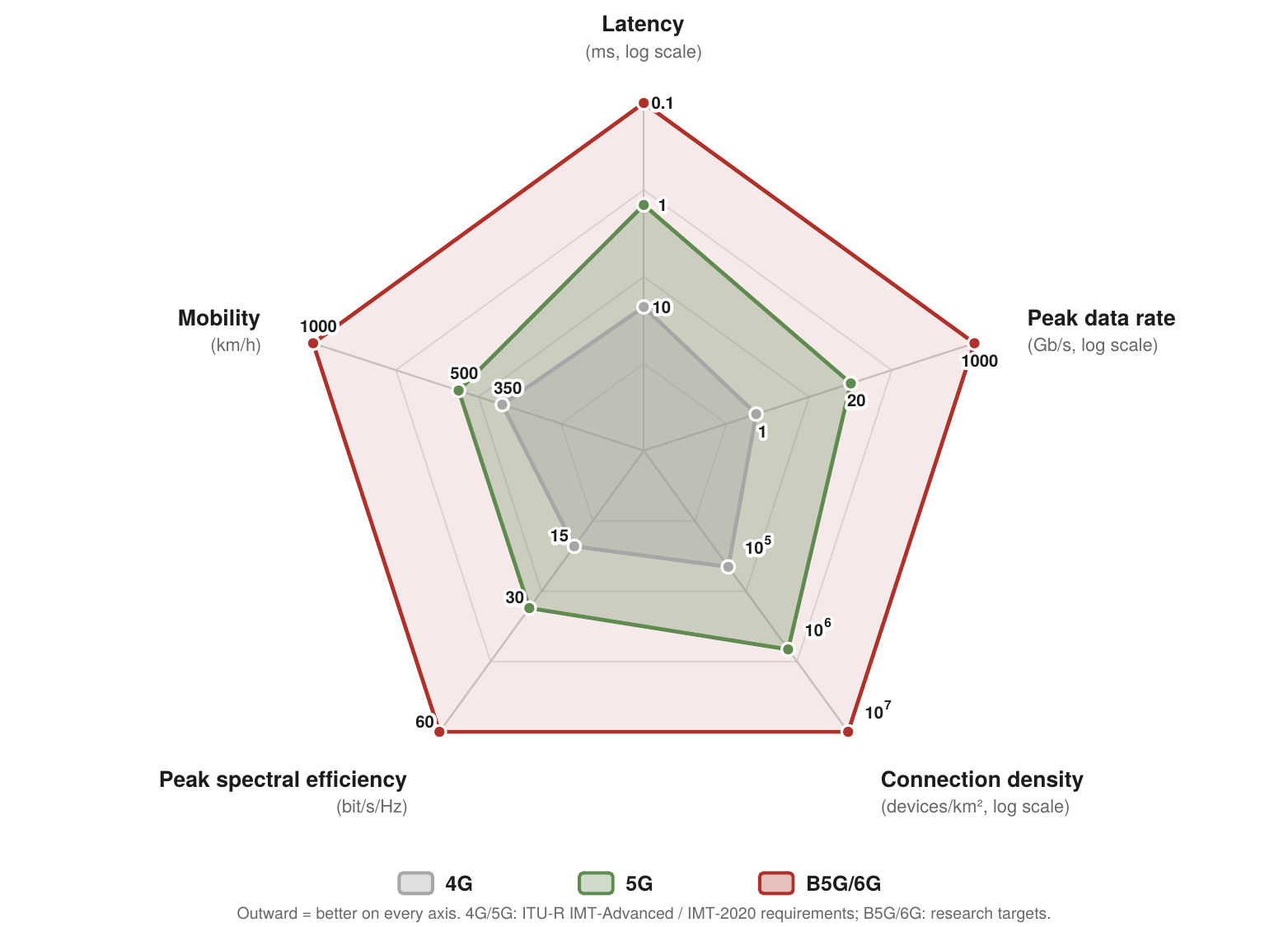}
    \caption{Comparison between 4G, 5G, and B5G/6G.}
    \label{5gVSb5g}
\end{figure}

Building on the innovations of 5G, the evolution toward Beyond‑5G and early 6G architectures is focused on extending and enhancing the capabilities of 5G by supporting emerging service classes such as ultra-Massive Machine-Type Communications (u-mMTC), extremely enhanced Mobile Broadband (eX‑eMBB), and 6G Ultra-Reliable Low-Latency Communications (6G‑URLLC), while sustaining sustainability, intelligence, and ubiquitous coverage \cite{khalid2021advanced} (See Fig. \ref{5gVSb5g} for a summary).

Key technical trends and architectural elements of B5G/6G include:
\begin{itemize}
    \item \textbf{Expansion into mmWave and THz Spectrum:} 
    Expansion into the mmWave and THz bands, coupled with (ultra)massive MIMO and highly directional beamforming, is being investigated for terabit-class peak physical layer data rates in B5G/6G; realizing such gains end-to-end will require new RF/antenna hardware and accurate channel/propagation models \cite{giordani2020toward}. Spectrum studies for IMT-2030 also include upper mid-band frequencies such as 7.125–8.4 GHz \cite{sampath2024imt2030}, so B5G/6G will not be confined to mmWave and THz bands.
    \item \textbf{Integration of AI/ML:} B5G embeds AI-native mechanisms, enabling real-time resource control, predictive network slicing, and self-healing capabilities; which means transitioning toward zero-trust security frameworks \cite{tariq2020speculative}.
    \item \textbf{Edge Computing and Distributed Intelligence:} Multi-access Edge Computing (MEC) are combined with Information-Centric Networking (ICN) to enable ultra-low latency while distributing data processing across dense, heterogeneous access points \cite{9845700}.
    \item \textbf{Integration with Non-Terrestrial Networks (NTNs):} Use of mmWave and THz bands, as well as satellite and aerial links introduces non-stationary, directional transmissions and multi-modal connectivity, which necessitates novel channel models for propagation and path loss \cite{3gppSatelliteComponents}.
    \item \textbf{Standardization and Roadmapping:} Standardization efforts for B5G are underway through various international bodies.
    \begin{itemize}
        \item 3GPP Release 18 (5G-Advanced) brings enhanced MIMO, RedCap for IoT, NTN support, and embedded AI/ML capabilities \cite{3gpp21916}.
        \item B5G/6G efforts, led by initiatives such as Japan’s NICT, ETSI’s RIS/THz/ENI working groups, and ITU's IMT‑2030 roadmaps, focus on defining architectures that blend terrestrial, satellite, and intelligent surface technologies \cite{mourad2020baseline, europaBeyond2024}.
    \end{itemize}
\end{itemize}

\section{Related Work} \label{sec:related-work}
In this section we start by listing and comparing our work to surveys on the same or closely related topic, then we give an overview about PNAs and their widespread exploitation. 
% To Do fill in the intro
\subsection{Recent Traffic Analysis Surveys} \label{subsec:surveys}

\begin{table*}[t]
\centering
\caption{Comparison of selected surveys on traffic analysis and privacy/security in wireless networks.}
\label{tab:survey_comparison}
\adjustbox{width=\textwidth}{
\begin{tabular}{c c c c c}

\hline
\textbf{Survey} & \textbf{Scope / Focus} & \textbf{Coverage of PNAs} & \textbf{Consider 5G} & \textbf{Consider B5G/6G} \\ \hline
\small

Rahbari \textit{et al.}~\cite{rahbari2015secrecy} & Passive/active SCA (wireless comms.) & General PNA concepts; not specific to cellular & No & No \\ \hline

Spreitzer \textit{et al.}~\cite{spreitzer2017systematic} & Tax. of mobile SCA (power, EM, network) & Mentions PNAs, not focus & No & No \\ \hline

Conti \textit{et al.}~\cite{conti2018dark} & Network TA for mobile users & Mostly Wi-Fi and AP/device assumptions & No & No \\ \hline

Salman \textit{et al.}~\cite{salman2020review,salman2021data} & ML-based TC & Indirectly; assumes access to TCP/IP stack & No & No \\ \hline

Kumar \textit{et al.}~\cite{kumar2021smartphone} & Smartphone TA & User/system behavior; malware focus & No & No \\ \hline

Papadogiannaki \& Ioannidis~\cite{papadogiannaki2021survey} & Encrypted TA (tech./counterm.) & Broad categories (website, app, OS, PII) & No & No \\ \hline

Cao \textit{et al.}~\cite{cao2019survey} & Security aspects of 5G & No focus on PNAs & Yes & No \\ \hline

Khan \textit{et al.}~\cite{khan2019survey} & Privacy thr. in 5G & High-level only; no PNAs & Yes & No \\ \hline

Sandeepa \textit{et al.}~\cite{Sandeepa_2022} & Privacy in B5G (id., loc., auth, regulation) & No PNA analysis & No & Yes \\ \hline

Ramezanpour \textit{et al.}~\cite{ramezanpour2022securityprivacyvulnerabilities5g6g} & 5G/6G and Wi-Fi 6 coexistence & Mentions SCAs; no feasibility analysis & Partial & Yes \\ \hline

Harvanek \textit{et al.}~\cite{s24175523} & Physical-layer security thr. (4G/5G) & Signal-level eavesdropping only & Yes & No \\ \hline
Wani \textit{et al.}~\cite{wani2024security} & 5G NSA vuln. (based on 4G attacks) & Focus on cataloging attacks; little on PNAs & Yes (NSA) & No \\ \hline

Saeed \textit{et al.}~\cite{saeed2025comprehensive} & 6G security thr. (layered taxonomy) & Mentions passive thr.; no PNA feasibility & No & Yes \\ \hline

Devi \textit{et al.}~\cite{DEVI2025100891} & Physical-layer security mechanisms (5G/6G) & Acknowledge eavesdropping, not PNAs & Yes & Yes \\ \hline

\textbf{Our work} & \textbf{Feasibility of PNAs in 5G/B5G} & \textbf{Explicit, detailed analysis} & \textbf{Yes} & \textbf{Yes} \\ \hline

\end{tabular}
}

\smallskip
\begin{footnotesize}
\begin{itemize}
    \item [] \textbf{Attacks:} SCA = Side-Channel Attacks, TA = Traffic Analysis, TC = Traffic Classification.
    \item [] \textbf{Abbreviations:} comms. = communications, cell. = cellular, thr. = threats, tax. = taxonomy, techn. = technique, counterm. = countermeasures, id. = identity, loc. = location, vuln. = vulnerabilities
    
\end{itemize}
\smallskip
\end{footnotesize}
\end{table*}

\begin{figure*}[ht]
    \centering
    \includegraphics[width=\textwidth]{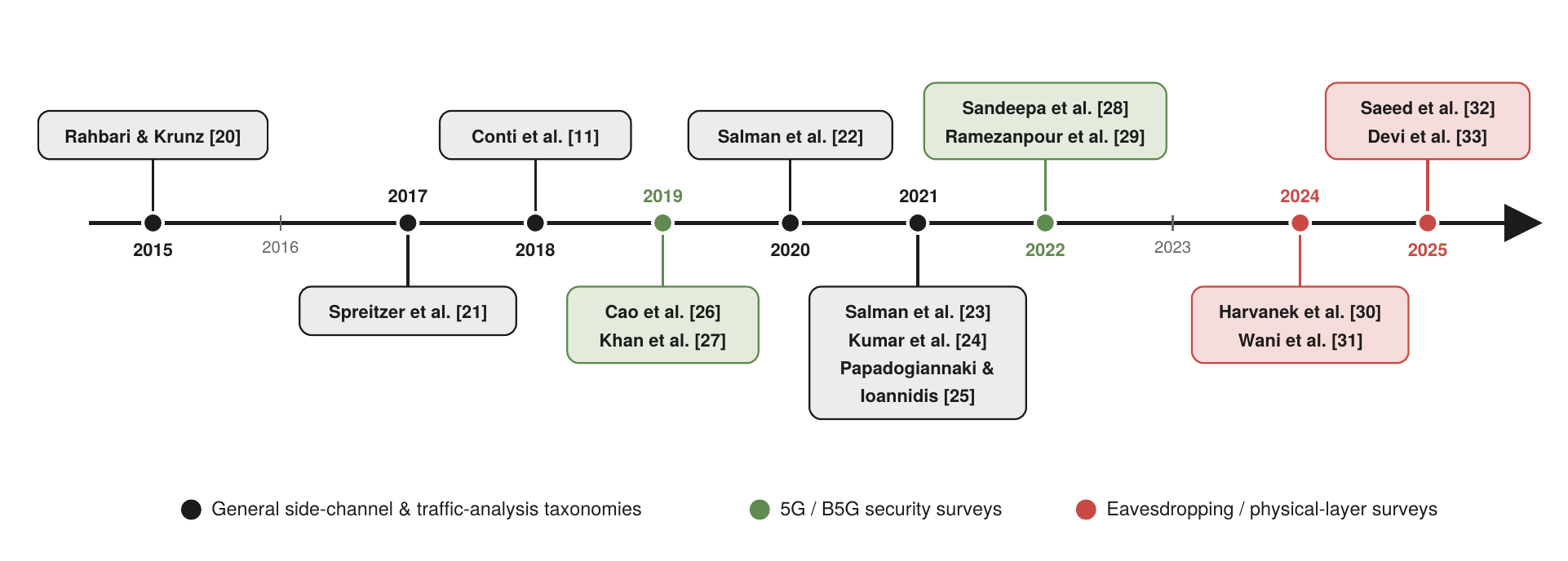}
    \caption{Publication year of selected key surveys covering PNAs.}
    \label{fig:timeline}
\end{figure*}

Several surveys on traffic analysis have been published in recent years (see Fig.~\ref{fig:timeline} for a timeline), and they fall into three broad clusters relative to our scope: those that classify side-channel or traffic-analysis techniques in general, without regard to the underlying wireless technology; those that survey 5G or B5G/6G security broadly, without focusing on passive attacks specifically; and those that touch on passive attacks or eavesdropping within a 5G/B5G context, but without assessing the practical feasibility of reproducing known attack techniques. None of the surveys in any of these three clusters examine the feasibility of adapting known passive attacks to 5G and B5G networks.

\subsubsection{General side-channel and traffic-analysis taxonomies} 
The earliest works in our comparison, Rahbari and Krunz~\cite{rahbari2015secrecy} and Spreitzer et al.~\cite{spreitzer2017systematic}, establish foundational taxonomies for side-channel attacks in wireless and mobile systems. Rahbari and Krunz~\cite{rahbari2015secrecy} describe passive traffic-analysis and active jamming attacks based on side-channel leakage, arguing that link-layer headers cannot be encrypted without prohibitive overhead, and advocate physical-layer security techniques to complement payload encryption; however, their treatment of MIMO systems predates 5G-specific radio characteristics such as massive MIMO beamforming. Spreitzer et al.~\cite{spreitzer2017systematic} extend this taxonomy across all side-channel modalities (power, electromagnetic, and network-based), classifying attacks by activity (passive/active) and invasiveness; because PNAs are only one branch of this broader taxonomy, cellular-specific constraints are not its focus. Conti et al.~\cite{conti2018dark} narrow the scope to network traffic analysis for mobile users specifically, categorizing 60 works by analysis goal, capture point, target platform, and encryption layer. Notably, only four of the works they survey~\cite{musa2012tracking,barbera2013signals,wang2015know,ruffing2016smartphone} assume a network-only capture point (as opposed to AP-side or on-device access), and all four target Wi-Fi, which, as discussed in Section \ref{subsec:challenges5g}, is considerably easier to eavesdrop passively than cellular communication. Salman et al.~\cite{salman2020review,salman2021data} and Kumar et al.~\cite{kumar2021smartphone} similarly assume, respectively, access to the full TCP/IP stack (with only application payload encrypted) and malware-focused device instrumentation, neither of which corresponds to a strictly external passive eavesdropper. Papadogiannaki and Ioannidis~\cite{papadogiannaki2021survey} offer the broadest of this cluster, cataloging traffic analysis by identification type (website, application, device, OS, PII, location), dataset, and technique, but again without considering the specific wireless access technology in use. Collectively, this cluster establishes the conceptual vocabulary this survey builds on (Section \ref{subsec:attack-types}), but leaves open exactly the question we address: which of these established techniques remain viable once the access technology is 5G or beyond.

\subsubsection{5G and B5G/6G security surveys}
A second cluster surveys 5G and B5G/6G security more broadly, without a specific focus on passive traffic analysis. Cao et al.~\cite{cao2019survey} and Khan et al.~\cite{khan2019survey} examine 5G security and privacy threats at a largely architectural and regulatory level, cataloging potential vulnerabilities without evaluating whether specific traffic-analysis techniques are practically reproducible. Sandeepa et al.~\cite{Sandeepa_2022} extend this to B5G, focusing on identity leakage, location privacy, authentication, and regulatory considerations, but likewise without a passive-attack feasibility analysis. Ramezanpour et al.~\cite{ramezanpour2022securityprivacyvulnerabilities5g6g} address the intersection of 5G/6G and Wi-Fi 6 coexistence and acknowledge side-channel risks, but do not assess whether attacks demonstrated on one technology transfer to the other.

\subsubsection{Physical-layer and threat-taxonomy surveys touching on eavesdropping}
A third cluster comes closest to our scope by explicitly discussing eavesdropping, but stops short of a feasibility analysis. Harvanek et al.~\cite{s24175523} survey physical-layer security threats in 4G and 5G comprehensively, including signal-level eavesdropping, jamming, spoofing, and rogue base station detection using SDR testbeds, but do not consider metadata-based passive attacks such as traffic fingerprinting. Wani et al.~\cite{wani2024security} survey 5G Non-Standalone (NSA) vulnerabilities by mapping known 4G attacks onto current NSA deployments and experimentally validate a small number of exploits, including an IMSI-leak tracking attack, on commercial smartphones; however, their scope is limited to the NSA architecture, and they note but do not deeply investigate metadata-analysis attacks like encrypted traffic fingerprinting. Saeed et al.~\cite{saeed2025comprehensive} present a layered taxonomy of 6G threats spanning physical, connection, and service layers, including passive eavesdropping, but only at a conceptual level without feasibility or SDR-based validation. Devi et al.~\cite{DEVI2025100891} similarly review physical-layer security countermeasures for 5G/6G (artificial noise, cooperative relaying, intelligent reflecting surfaces) and acknowledge eavesdropping as a motivating risk, without analyzing practical PNAs or side-channel exploitation.

\medskip
Table~\ref{tab:survey_comparison} summarizes this comparison across scope, PNA coverage, and 5G/B5G consideration. The pattern across all three clusters is consistent: surveys either treat traffic analysis generically without regard to access technology, or treat 5G/B5G security broadly without a feasibility-oriented treatment of passive attacks specifically. Our work is, to our knowledge, the first to combine both dimensions explicitly, offering a systematic feasibility assessment of previously demonstrated passive attacks under 5G and B5G radio and protocol constraints.

\subsection{Overview of Passive Network Attacks} \label{subsec:passive_attacks}

Packet trace analysis has been used in recent years to demonstrate how PNAs can infer detailed user-related information by eavesdropping encrypted communications. These attacks can reveal the user's operating system and browser version, websites visited, running applications, and even the specific video content being streamed. Such analyses rely on metadata from captured network traces, typically encrypted, making the attacks difficult to detect. These traces can be collected at different layers of the protocol stack (e.g., data-link, network, or application layer) and from various network vantage points (e.g., via Wi-Fi eavesdropping, Tor\footnote{Tor (The Onion Router) is a low-latency anonymity network that routes user traffic through multiple volunteer-operated relays, encrypting it in layers (“onion routing”) to conceal the user’s IP address and communication patterns from network observers.} networks, or device-level logging after privilege escalation). In all cases, these attacks highlight the inherent leakage in side-channel information even when payload data is fully protected.

\subsubsection{Network Traffic Analysis and Passive Network Attacks}

Packet trace analysis is a class of side-channel attacks~\cite{spreitzer2017systematic} that exploits metadata (such as packet sizes, arrival times, and flow direction) to infer sensitive information, even when traffic is encrypted at the transport or application layer (e.g., via HTTPS). PNAs rely entirely on metadata without requiring interaction with or modification of packets.

Classic ML methods such as k-Nearest Neighbors (kNN), Support Vector Machines (SVM), and Random Forests (RF) have been widely applied. For example, Reed \textit{et al.}~\cite{reed2016leaky,reed2017identifying} demonstrated high-accuracy video fingerprinting over encrypted DASH traffic using kd-tree classifiers, including for Netflix streams. Taylor \textit{et al.}~\cite{taylor2016appscanner,taylor2017robust} proposed AppScanner to identify 110 Android apps over encrypted HTTPS traffic with up to 99\% accuracy, while Wang \textit{et al.}~\cite{wang2015know} used RF classifiers to identify smartphone applications. More recently, deep learning (DL) approaches (including CNNs, LSTMs, and hybrid architectures) have achieved similar or higher accuracy for mobile and encrypted traffic classification~\cite{aceto2019mobile,aceto2020toward,rezaei2019large,d2021network}. These studies generally assume access to flows with bidirectional metadata.

Some works consider packet-level sequences with minimal visibility. For instance, Acar \textit{et al.}~\cite{acar2020peek} inferred smart home device actions from Wi-Fi, ZigBee, and BLE traffic using only packet timing and size, while Meneghello \textit{et al.}~\cite{meneghello2020smartphone} applied sequence-to-sequence learning over LTE PDCCH traffic for smartphone identification. Shapira \textit{et al.}~\cite{shapira2019flowpic} and Montieri \textit{et al.}~\cite{montieri2021packet} explored unsupervised and DL-based traffic classification at the flow or packet level.

Key types of PNAs demonstrated across networks include:

\begin{itemize}
    \item \textbf{Website Fingerprinting (WF):} Rimmer \textit{et al.}~\cite{rimmer2018automated} achieved 96\% accuracy in classifying visited websites over Tor using DL architectures such as stacked autoencoders, CNNs, and LSTMs. Juarez \textit{et al.}~\cite{juarez2014critical} highlighted overfitting risks and practical challenges under real-world conditions. Most WF studies target Wi-Fi or VPN-encrypted traffic, with limited application to cellular networks.
    \item \textbf{Video and App Fingerprinting (VF/AF):} Reed \textit{et al.}~\cite{reed2016leaky,reed2017identifying} and Dubin \textit{et al.}~\cite{dubin2017know} used kd-tree or SVM classifiers to identify encrypted video streams or smartphone applications. Petagna \textit{et al.}~\cite{petagna2019peel} extended this to Tor, demonstrating app identification despite multi-hop encryption.
    \item \textbf{Smart Home and Behavioral Inference:} Acar \textit{et al.}~\cite{acar2020peek} and Aiolli \textit{et al.}~\cite{aiolli2019mind} inferred device actions and cryptocurrency wallet activity from encrypted traffic. Conti \textit{et al.}~\cite{conti2015can} and Muehlstein \textit{et al.}~\cite{muehlstein2017analyzing} analyzed Android traffic to deduce user behavior and app usage patterns.
    \item \textbf{De-anonymization in Privacy-Enhancing Networks:} Karunanayake \textit{et al.}~\cite{karunanayake2021anonymisation} surveyed passive attacks on networks like Tor, including fingerprinting, timing correlation, and traffic confirmation attacks, highlighting that anonymizing overlays do not fully prevent side-channel leaks.
\end{itemize}

\subsubsection{Passive Attacks on Cellular Networks} \label{subsubsec:pna-cell}

While passive attacks are well-documented in Wi-Fi and anonymizing networks, their extension to cellular networks, including LTE, 3G, and 4G, has also been demonstrated with varying degrees of success.
Stöber \textit{et al}~\cite{stober2013you} showed that even without payload access, user identification is possible over 3G and LTE networks by analyzing traffic patterns. Their work relied on timing and volume features, enabling fingerprinting of smartphone users based solely on metadata. Similarly, Meneghello \textit{et al}~\cite{meneghello2020smartphone} applied DL (1D-CNN) to LTE’s control channel data (PDCCH) and demonstrated accurate fingerprinting of specific smartphones using only physical layer side-channel features.
Website fingerprinting over cellular links has also been shown to be effective. Kohls \textit{et al}~\cite{kohls2019lost} and Rupprecht \textit{et al}~\cite{rupprecht2019breaking} performed passive Layer 2 traffic analysis in LTE networks and successfully inferred visited websites by observing encrypted traffic in the RLC (Radio Link Control\footnote{The Radio Link Control layer handles segmentation, reassembly, error correction, and reliable delivery of data between the UE and the base station in LTE and 5G NR.}) and PDCP (Packet Data Convergence Protocol\footnote{The PDCP layer provides header compression, encryption, integrity protection, and in-order delivery of user-plane and control-plane data in LTE and 5G NR.}) layers. These studies highlight that despite encryption at upper layers, the structure and scheduling behavior of lower layer traffic reveals enough to compromise user privacy.
Khanna \textit{et al}~\cite{khanna2015remote} proposed remote device fingerprinting techniques based on passive measurement of MAC/IP behavior, clock skew, and network layer metadata. Though not limited to LTE, their methods are particularly effective in mobile networks where device-specific timing characteristics can be measured due to frequent reconnections or network handovers.
Trinh \textit{et al}~\cite{trinh2020mobile} introduced an approach using LTE control channel features to classify the applications and services accessed by a device. Using DL models, they demonstrated that even control plane metadata, such as resource allocation and scheduling information, can be exploited to infer sensitive information without touching the encrypted payload.

These LTE/4G-specific attacks serve as important baselines for understanding what passive analysis can achieve in mobile environments. However, extending these attacks to 5G introduces several new challenges.

Extending these attacks to 5G changes the conditions under which they operate rather than introducing entirely new protection mechanisms: LTE already applies PDCP ciphering to user-plane traffic \cite{3gpp36323} and supports multi-antenna transmission, including beamformed data transmission \cite{3gpp36213}. What differs in NR is the degree to which beamforming is integrated into the radio procedures themselves. NR builds beam management into initial access, through beam-swept synchronization signal blocks and beam-specific control resource sets, and supports beamformed transmission of the control channel \cite{3gpp38213}, whereas the legacy LTE PDCCH is typically transmitted across the whole cell or sector. For an eavesdropper, the relevant question is therefore not whether it is exactly aligned with a given beam, but whether it receives the selected transmission with sufficient SINR to decode its format, whether through the main lobe, a side lobe, or a multipath component.
Furthermore, 5G traffic patterns are more dynamic, fragmented, and often multiplexed across services, making side-channel separation more complex. Norrman \textit{et al}~\cite{norrman2016protecting} and Khan \textit{et al}~\cite{khan2018defeating} studied user privacy threats from rogue base stations and demonstrated possible identity exposure during 5G synchronization. 
Capturing usable traces in 5G typically requires SDRs and specialized demodulation tools. Works like Wei et \textit{et al}~\cite{wei2016software}, Duarte et \textit{et al}~\cite{duarte2019software}, and Vo-Huu et \textit{et al}~\cite{vo2016fingerprinting} explored SDR-based fingerprinting and signal extraction methods, but 5G's physical layer complexity presents practical limitations. 
Additionally, concurrent application traffic causes flow interleaving, making separation and classification harder, especially under PDCP-layer ciphering. Classifiers often assume clean data; an unrealistic expectation in 5G’s dynamic environment. Also, traffic varies significantly across devices, complicating generalization. Aceto \textit{et al}~\cite{aceto2019mobile} and Ismailaj \textit{et al}~\cite{ismailaj2021deep} note that even state-of-the-art DL models underperform on noisy, mixed data or new devices.
Nevertheless, increased deployment density of 5G small cells may inadvertently expose users to attackers within physical proximity, potentially making some forms of passive monitoring easier than in previous generations.

\paragraph*{Remarks} The diversity and increasing sophistication of PNAs over encrypted traffic raise serious privacy concerns. While many of these attacks have been demonstrated successfully in Wi-Fi and LTE environments, their feasibility in 5G/B5G remains largely unexplored. This survey aims to evaluate to what extent known passive attacks, ranging from system fingerprinting to behavioral and location inference, can be realistically reproduced in 5G/B5G networks. However, performing such analysis in 5G/B5G poses significant challenges: encrypted link-layer traffic, spatially selective communication, limited availability of SDR tools, and the difficulty of acquiring clean, labeled data. These issues will be explored in depth in the following sections, where we assess the characteristics of 5G/B5G and the practical feasibility of each type of attack.

\section{Methodology} \label{sec:methodology}

We present in this section the data collection methodology and survey various works about PNAs in Wi-Fi, LTE, and cellular communications.

\subsection{Data selection criteria} \label{subsec:data-selection}
To assess the practical feasibility of PNAs in 5G and B5G environments, we adopted a targeted selection strategy rather than an exhaustive enumeration of all existing attack literature. Our primary objective is not to catalog every theoretical variation of traffic analysis, but to identify distinct classes of attack vectors that have been proven effective in previous generations (Wi-Fi, LTE) and rigorously evaluate their reproducibility under 5G physical and protocol constraints.

We screened literature published over the last 14 years (2013–2026) to capture the evolution of attacks from the early 4G era to current 5G and early 6G deployments. From an initial pool of hundreds of studies, we filtered for works that met the following inclusion criteria:
\begin{itemize}
    \item \textbf{Peer-Reviewed Validation:} We primarily considered works published in reputable journals and conferences. We additionally included a small number of recent preprints (\cite{cheng2023watching, li2026robust, ludant2024unprotected}) that report experimental results on live LTE/5G systems directly relevant to our feasibility analysis.
    \item \textbf{Impact and Relevance:} We selected the top-cited papers per year that introduced novel methodologies (e.g., the first instance of video fingerprinting using variable bit rate, or the first deep learning-based website fingerprinting attack).
    \item \textbf{Reproducibility:} Priority was given to studies providing clear threat models and experimental setups (e.g., specific SDR hardware or dataset characteristics) essential for feasibility analysis.
    \item \textbf{Threat Model:} We focus in this study on two main types of PNAs. We investigated: (i) packet trace analysis based on captured information, and (ii) passive localization or tracking of users.
\end{itemize}

This process resulted in a core set of \textbf{44 seminal works}. This dataset represents the "state-of-the-art" in attack sophistication. By focusing on these high-impact exemplars, we can perform a deep-dive technical analysis of why specific mechanisms (e.g., packet size side-channels) succeed or fail in the face of 5G beamforming and encryption, providing a qualitative depth that a broader quantitative survey would lack.

\subsection{Collected data} \label{subsec:data}

We detail below 44 selected works that form the basis of our analysis (see Table \ref{tab:pna-summary} for an overview):

\setlength{\tabcolsep}{3pt} 
\renewcommand{\arraystretch}{0.9} 
\begin{table*}[t]
    \centering
    \caption{Summary of passive network attacks selected in both user/traffic identification (upper half) and position tracking (lower half).}

    \begin{tabular}{c c c c c c c c}
        \hline
        \textbf{Year} & \textbf{Authors [ref.]} & \textbf{Outcome} & \textbf{Channel used} & \textbf{Protocol(s) / tool} & \textbf{Technique/ML used } & \textbf{Scale} & \textbf{Accuracy} \\
        \hline
        \small 
        
        2013 & Barbera \textit{et al.} \cite{barbera2013signals} & Social rlt. ID & Link layer & Wi-Fi & Correlations & 460 users & N/A \\

        2013 & Stöber~\cite{stober2013you} & User ID & Link-layer & 3G LTE & kNN, SVM & 20 users & 90\% \\
        
        2014 & Chen \textit{et al.} \cite{chen2014fingerprinting} & OS ID & Cellular uplink (L3) & Mobile ISP & Signature + DT & 2 mob. op. & 95\% \\
        
        2015 & Wang \textit{et al.}~\cite{wang2015know} & Application ID & Link-layer & 802.11a/b/g & RF & 20 apps & 94\% \\
        
        2016 & Ruffing \textit{et al.} \cite{ruffing2016smartphone} & OS ID & WiFi (L2) & 802.11 frames & Statistical & 4 mob. plat. & ~95\% \\
         
        2016 & Saltaformaggio \textit{et al.} \cite{saltaformaggio2016eavesdropping} & Activity ID & WiFi/LTE (L3) & Encrypted traffic & Statistical + Heuristic & 35 app act. & 78\% \\
        
        2016 & Taylor \textit{et al.}~\cite{taylor2016appscanner,taylor2017robust} & Application ID & Network-layer & HTTPS/TLS & SVM (ML) & 110 apps & 99\%\\ %up to 
        
        2016 & Reed \textit{et al.}~\cite{reed2016leaky} & Video ID & Link-layer & DASH & kd-tree & 25 videos & 90\%\\
        
        2017 & Reed \textit{et al.}~\cite{reed2017identifying} & Video ID & Network-layer & DASH (Netflix) & kd-tree & 42k videos & 99.5\%\\
        
        2017 & Dubin \textit{et al.}~\cite{dubin2017know} & Video ID & Network-layer & Youtube & 1-NN, SVM & 15k videos & 95\%\\

        2017 & Muehlstein \textit{et al.}~\cite{muehlstein2017analyzing} & OS, browser, app. ID & Transport-layer & HTTPS/SSL & SVM-RBF & 144 labels & 96\% \\
        
        2017 & Rimmer \textit{et al.}~\cite{rimmer2018automated} &  Website ID & Network layer & ToR & SAE, CNN, LSTM & 100 websites & 96\%\\
        
        2019 & Aceto \textit{et al.}~\cite{aceto2019mobile} & Traffic ID & All L7-layers & All & DL & 45 apps & 83-93\% \\
        
        2019 & Petagna \textit{et al.}~\cite{petagna2019peel} & Android apps ID & Transport Layer & TCP (ToR) & k-NN, RF, SVM & 10 apps & 97\% \\

        2019 & Shapira \textit{et al.}~\cite{shapira2019flowpic} & Traffic ID & Flows (L4) & VPN / ToR & CNN & 7 types & 68-98\% \\
        
        2019 & Rezaei \textit{et al.}~\cite{rezaei2019large} & Mobile apps ID & Flows (L4) & UDP, TCP, HTTP(S) & CNN+LSTM & 80 mobile apps & 95\% \\

        2019 & D’Angelo \textit{et al.}~\cite{d2021network} & Traffic ID & Flows (L4) & 12+ protocols & CNN/LSTM+SAE+NN & 4 categories & 99\% \\
        
        2020 & Acar \textit{et al.}~\cite{acar2020peek} & SH activities ID & Link-layer & WiFi, ZigBee, BLE & kNN, RF, HMM, +a & 22 devices  & 88-100\%\\
        
        2020 & Meneghello \textit{et al.}~\cite{meneghello2020smartphone} & Smartphone ID & LTE (L1) & PDCCH & 1D-CNN & 40 sim. users & 75-90\% \\
        
        2020 & Aceto \textit{et al.}~\cite{aceto2020toward} & Traffic ID & All L7-layers & All & DL & 45 apps & 83-93\% \\

        2020 & Trinh \textit{et al.}~\cite{trinh2020mobile} & App/Service ID & LTE (L2) & PDCCH & RNN, CNN, MLP, +a & 6 apps/3 serv. & 98\% \\
        
        2020 & Wang \textit{et al.}~\cite{wang2020automatic} & Application ID & Biflows (L4) & TLS & RNN+CNN & 80 apps & 93-99\% \\

        2020 & Gijon \textit{et al.}~\cite{gijon2020encrypted} &  Traffic ID & LTE (L4) & CTR & AHC & 8 labels & N/A \\
        
        2021 & Montieri \textit{et al.}~\cite{montieri2021packet} &  Pkt. ID & Biflows (L4) & TCP & CNN, RNN, CompNN  & 16 labels & N/A \\

        % this is a post-processing basically to improve performance of traffic classification
        2021 & Zhao \textit{et al.}~\cite{zhao2021optimized} & Traffic ID & Transport Layer & N/A & K-means & 11 labels & 88\% \\

       2023 & Cheng \textit{et al.}~\cite{cheng2023watching} & VoLTE ID & LTE/5G (L2/L3) & srsRAN & Eavesdropping & 60h traffic & 100\% \\

       2023 & Budykho \textit{et al.}~\cite{budykho2023fine} &  Entity ID & 5G RRC (L3) & TrackDev~\cite{githubGitHubFmsectrackdev} & Trackability analysis & $\approx 10-20$ users & N/A \\

       2023 & Xiong \textit{et al.}~\cite{xiong20235g} & User act. ID & 6G UAV (L3) & GTP/IP over 6G & CNN & Small testbed & N/A \\

       2023 & Björklund \textit{et al.}~\cite{10060390} & Video ID & L3-L4 & TLS over TCP/TLS & k-d tree & 1000+ videos & 99\% \\

       2024 & Wan \textit{et al.}~\cite{wan2024nr} & RAN telemetry ID & 5G NR (L1) & PDCCH/DCI & Decoding RRC & 3 5G SA RAN & 99\% \\

       2024 & Wani \textit{et al.}~\cite{wani2024security} & IMSI ID & 5G NSA (L1–L3) & LTE RRC / NAS & Passive sniffing & 8 smartphones & N/A \\

       2024 & Marañón \textit{et al.}~\cite{10823417} &  Enc. app ID & 5G bursts (L3-L4) & Passive pkt. capture & kNN, RF, LSTM & 8 mobile apps & $\approx85\%$ \\

       2025 & Jawne \textit{et al.}~\cite{jawne2025ai} & 5G devices ID & 5G NR RF (L1) & SDR testbed & ResNet & 4 UEs & $\approx$100\% \\

       2025 & Zhang \textit{et al.}~\cite{zhang2025passive} & User act. ID & PUCCH (L1) & 3GPP 5G NR & RF, SVM, k-NN & 5 smartphones & $\approx$100\% \\

       2025 & Björklund \textit{et al.}~\cite{usenixEndangeredPrivacy} & Video ID & Video streams (L3-L4) & VPN/Wi-Fi & k-d tree & 240,000 videos & 99.5\% \\

       2026 & Li \textit{et al.}~\cite{li2026robust} & Application ID & 5G NR uplink (L1) & PUSCH/PUCCH & DL (cross-modal) & 78 apps & $>90$\% \\

        2026 & Wang \textit{et al.}~\cite{wang2026app} & Application ID & LTE/5G (L2, relay) & PDCP & ML classifier & 40 apps & $99.02$\% \\
        
        \hline
        \hline
        
        2015 & Ateniese \textit{et al.} \cite{ateniese2015no} & Location TR & WiFi/Cellular (L3) & TLS & RF, k-NN & 100+ locations & 90-95\% \\

        2016 & Yang \textit{et al.}~\cite{yang2016passive} & Passive TR & Link-layer & Wifi & 1-NN & 5 test points & N/A \\

        2019 & Kohls \textit{et al.}~\cite{kohls2019lost} & User ID, TR & LTE/4G (L2) & RLC/PDCP & k-NN, & 50 websites & 92-95\% \\
        
        2019 & Rupprecht \textit{et al.}~\cite{rupprecht2019breaking} & Website ID, TR & LTE/4G (L2) & RLC/PDCP & k-NN & 50 websites & 89\% \\

       2022 & Kotuliak \textit{et al.}~\cite{kotuliak2022ltrack} &  UE TR & LTE (L1) & SDR sniffer & Time measurement & 17 phones & 90\% \\

       2023 & Ludant \textit{et al.}~\cite{ludant20235g} &  User pres. TR & 5G (L1) & PDCCH/DCI & DCI decoding & Single 5G cell & 94\% \\

       2024 & Ludant \textit{et al.}~\cite{ludant2024unprotected} &  User TR & MAC/PHY (L1-L2) & MAC CE & Signal analysis & COTS UEs & 96\% \\
       
        \hline
    \end{tabular}
    
    \smallskip
    \begin{footnotesize}
    \begin{itemize}
        \item[] \textbf{Problems}: ID = identification, TR= Tracking, SH = Smart home, enc. = encrypted, rlt. = relationship, sim. users = simulated users, pkt.= packet, app act. = application activities, mob. op. = mobile operator, mob. plat. = mobile platform, user pres. = user presence, UE = user equipment, IMSI = International Mobile Subscriber Identifier.
        \item[] \textbf{Machine Learning}: kNN = $k$ Nearest Neighbors, SVM = Support Vector Machine, RBF = Radial Basis Function, DT = Decision Trees, RF = Random Forests, HMM = Hidden Markov Model, AHC = Agglomerative Hierarchical Clustering.
        \item[] \textbf{Deep Learning}: MLP = Multilayer Perceptron, RNNs = Recurrent Neural Networks, CNN = 1D- and 2D- Convolutional Neural Network, DNN = Deep Neural Network, LSTM = Long Short-Term Memory, AE = AutoEncoder, SAE = Stacked AutoEncoder, SDAE = Stacked Denoising Autoencoder, bi-GRU = Bidirectional Gated Recurrent Unit, CompNN = Composite Neural Networks, ResNet = Residual Neural Network.
        \item[] \textbf{Radios \& Networks}: RF = Radio Frequencies, RLC = Radio Link Control, RRC = Radio Resource Control, RAN = Radio Access Network, PDCP = Packet Data Convergence Protocol, PUSCH = Physical Uplink Shared Channel, PUCCH = Physical Uplink Control Channel, PDCCH = Physical Downlink Control Channel, MAC CE = Medium Access Control Control Element, CTR = Cell Traffic Recording, DCI = Downlink Control Information, VoLTE = Voice over LTE, GTP = GPRS Tunneling Protocol, 3GPP = The 3rd Generation Partnership Project.
        \item[] \textbf{Others}: N/A = Not Applicable, COTS = Commercial Off-The-Shelf.
    \end{itemize}
    +a = XGBoost, Adaboost, Random Forest, SVM, with RBF kernel, kNN, Logistic Regression, Naïve Bayes, and Decision Tree \\
    \end{footnotesize}
    
    \smallskip
    \label{tab:pna-summary}
\end{table*}

\subsubsection{About users' and traffic identification} Many works in Table \ref{tab:pna-summary} target user or application identification through traffic fingerprinting, although they differ widely in the data sources and platforms targeted. 

Barbera \textit{et al.}~\cite{barbera2013signals} collected Wi-Fi probe request frames (a type of management frame) that are unencrypted, using commodity NICs, and inferred social relationships by detecting overlapping SSIDs in the devices’ preferred network lists. Along the same line, Wang \textit{et al.}~\cite{wang2015know} show how to identify mobile applications over 802.11 by leveraging link‐layer side channels, namely packet size distributions and interarrival timing features, even when the payload is encrypted, while Ruffing \textit{et al.}~\cite{ruffing2016smartphone} proposed a passive OS identification method that infers a smartphone’s operating system from encrypted network traffic using spectral analysis of packet flows (frequency-domain features).

Other works relied on video and multimedia fingerprinting. Reed \textit{et al.}~\cite{reed2016leaky} introduced one of the earliest approaches for encrypted DASH streams: they infer the sequence of video segment sizes from observed packet flows and index sliding windows of segments using a kd‐tree classifier (based on a 6-dimensional key capturing total size and relative allocation across subintervals). This approach assumes relatively clean captures where timing and packet sizes can be resolved to reconstruct segment boundaries. Their later work~\cite{reed2017identifying} extends this method to larger catalogs and refines the kd-tree search and filtering steps, but at the cost of increased sensitivity to capture noise, packet loss, or incomplete traces. More recently, Björklund \textit{et al}~\cite{10060390} adapted the burst-fingerprinting approach to TLS-encrypted DASH streams by mapping observed packet bursts to segment size fingerprints and matching them via kd-tree search; their method appears to be the fastest video identification to date (above 98\% accuracy within 15 seconds of packet capture). Björklund and Duvignau~\cite{usenixEndangeredPrivacy} later extended this line of work into a protocol-agnostic, large-scale attack that identifies streaming content solely from timing and burst patterns, achieving high accuracy ($>$99.5\%) even on large catalogs (over 200k videos).

Some studies focused on user activity inference. Saltaformaggio \textit{et al.}~\cite{saltaformaggio2016eavesdropping} proposed NetScope, a passive eavesdropping framework that infers fine-grained in-app user activities by analyzing only packet metadata (sizes, timing, direction) and heuristics over burst/flow patterns. Even without access to payload, they achieve $\approx$78\% precision and $\approx$76\% recall over 35 app activities in their evaluation. In parallel, Stöber~\cite{stober2013you} showed that individual smartphones can be fingerprinted by passively observing background traffic patterns generated by installed applications (synchronization, updates) and exploiting side-channel features like periodicity, timing, and volume. Their method works purely via external network observation (without needing instrumented client logs). Chen \textit{et al.}~\cite{chen2014fingerprinting} presented a method for passive OS fingerprinting using flow metadata (e.g. TCP/IP header features) from an ISP or mobile-operator vantage point. Their approach uses standard flow-analysis features and classifiers on upstream traffic to distinguish operating systems and also detect tethering.

Other works emphasized application-level identification. Taylor \textit{et al.}~\cite{taylor2016appscanner}’s \textit{AppScanner} used TLS/HTTPS metadata (packet sizes, directions, timing) combined with aggregate session features (e.g. bytes, packet counts, durations) to fingerprint smartphone apps from encrypted traffic. Operating over captures or ISP/flow logs, it achieves high accuracy across a large set of apps. Similarly, Rimmer \textit{et al.}~\cite{rimmer2018automated} proposed a deep learning–based website fingerprinting attack over Tor, which learns implicit traffic features (size, timing, direction) without manual feature engineering. Though it assumes the ability to partition and label flows for training, their approach achieves high accuracy ($>$96\%) by leveraging large datasets and neural architectures. Muehlstein \textit{et al.}~\cite{muehlstein2017analyzing} showed that a passive adversary observing encrypted TLS/HTTPS traffic can infer the user’s operating system, browser, and application by exploiting packet ordering, sizes, timing, and TLS handshake metadata. Aceto \textit{et al}~\cite{aceto2019mobile, aceto2020toward} extended this line of work by applying DL models to encrypted mobile flows: they classify mobile app traffic in WiFi / LTE settings using full application-level flows (e.g. packet sequences, timing, payload lengths) as input. Their approach uses standard capture or flow logs (i.e. commodity collection setups) and leverages automatic feature extraction via neural networks. Along the same direction, Rezaei \textit{et al.}~\cite{rezaei2019large} developed a CNN + LSTM model for mobile app identification by combining per-flow features (e.g. packet sizes, timing, header/payload bytes in the first few packets) with sequential context from adjacent flows. Their approach assumes visibility of flow-level metadata and some handshake bytes, and is able to improve classification accuracy, especially on ambiguous flows. D’Angelo \textit{et al}~\cite{d2021network} broadened traffic classification to multiprotocol settings (including encrypted traffic) by training deep convolutional recurrent autoencoders on bidirectional flows to automatically extract spatio-temporal features, while Shapira \textit{et al.}~\cite{shapira2019flowpic} propose FlowPic, a method to convert encrypted VPN/Tor flows into 2D images (packet size vs time), and apply a CNN to classify traffic categories and applications. In experiments on ISCX VPN/Tor datasets, they achieve up to $\approx$99.7\% accuracy on category classification, though performance degrades for application classification under Tor. Petagna \textit{et al.}~\cite{petagna2019peel} also performed app deanonymization over Tor on Android devices by passively reconstructing TCP/Tor flows, extracting timing, burst, and size features, and classifying them via machine learning. Montieri \textit{et al.}~\cite{montieri2021packet} pushed traffic analysis to the packet level, using multitask deep learning to jointly predict parameters like packet direction, payload length, and inter-arrival time. Their method is evaluated on real mobile app datasets (MIRAGE), outperforms traditional baselines (Markov, Random Forest), and highlights that no single architecture suits all apps. Marañón and Duvignau~\cite{10823417} examined the feasibility of app identification from encrypted 5G packet traces, using only traffic-pattern features (timing, size, direction). They applied classical ML (k-NN, Random Forest) and LSTM to highlight that even in 5G environments, encrypted traffic remains vulnerable to pattern leakage.

Other investigations ventured into cross-technology and control-plane contexts. Acar \textit{et al.}~\cite{acar2020peek} designed a multi-stage passive privacy attack across Wi-Fi, ZigBee, and BLE networks. From purely sniffed MAC-layer timing and traffic-volume metadata, they successively infer device identity, state transitions, device state, and finally user activities. Meneghello \textit{et al.}~\cite{meneghello2020smartphone} and Trinh \textit{et al.}~\cite{trinh2020mobile} applied fingerprinting techniques to LTE control-plane side channels (i.e. PDCCH/DCI metadata). The former uses PDCCH signal observations and a 1D-CNN model for device or state fingerprinting, while the latter decodes DCI messages and builds a CNN classifier to classify apps and services running on user equipment solely from control-plane fingerprints, even though data traffic remains encrypted. Wang \textit{et al.}~\cite{wang2020automatic} also used a hybrid CNN + LSTM architecture to identify mobile apps from encrypted traffic (TLS flows), fusing byte-level and sequence-level features, and assuming access to flow-level metadata (not full payloads). Moreover, Gijon \textit{et al.}~\cite{gijon2020encrypted} and Zhao \textit{et al.}~\cite{zhao2021optimized} pursue unsupervised traffic classification: the former clusters radio-trace descriptors into service classes, while the latter fuses SOM with K-means to cluster flow-level transport features from the Moore dataset. Both require flow/connection semantics but operate without labels. Cheng \textit{et al.}~\cite{cheng2023watching} further demonstrated that a mobile-relay setup (e.g. built using srsRAN) can eavesdrop on VoLTE/VoNR in LTE/5G networks and recover transport-adjacent metadata such as call timing, duration, and direction, even when the traffic is encrypted, though without recovering voice content. Building on this same mobile-relay methodology, Wang \textit{et al.}~\cite{wang2026app} presented \emph{What-App?}, an app-usage fingerprinting attack that operates via a MITM relay in an operational cellular network rather than through over-the-air capture: because the relay decrypts PDCP traffic in transit before re-encrypting it toward the legitimate base station, the attacker obtains genuine plaintext flow-level metadata rather than an inferred side-channel. Their attack identifies 40 applications with up to 99.02\% accuracy and demonstrates, notably, cross-device and cross-network transferability, directly addressing the device/vendor generalization limitation this survey highlights in Section \ref{subsec:challenges5g}.

Finally, several works explored 5G-specific or physical-layer side-channels. Budykho \textit{et al.}~\cite{budykho2023fine} probed fine-grained trackability in 5G by analyzing RRC (Radio Resource Control) signaling traces with their TrackDev framework, showing how execution patterns in the control plane can link sessions to the same user/device, while Wan \textit{et al.}~\cite{wan2024nr} developed NR-Scope, a 5G SA telemetry tool that decodes unencrypted PDCCH/DCI control messages to extract fine-grained RAN scheduling and resource allocation information, enabling passive recovery of network-level telemetry without access to user payloads. Jawne \textit{et al.}~\cite{jawne2025ai} showed that wideband SDR-derived spectrograms, coupled with deep learning models (ResNet, Transformer, LSTM), can fingerprint devices at the RF layer in 5G, enabling identity recovery despite encryption of higher-layer protocols. Also, Zhang \textit{et al.}~\cite{zhang2025passive} proposed PTTF, a passive traffic analysis attack exploiting uplink HARQ (Hybrid ARQ) ACK/NACK patterns on the PUCCH in 5G NR/LTE. By extracting statistical time–frequency features from ACK/NACK power distributions, they applied Random Forest to achieve fine-grained App, category, and service identification without demodulating signals or cooperating with operators. Wani \textit{et al.}~\cite{wani2024security} systematically analyzed 5G NSA vulnerabilities and showed that, due to LTE anchoring, the system remains exposed to IMSI leakage attacks during attach procedures. This makes the attack architecture-specific, affecting NSA but not standalone 5G. At the experimental frontier, Xiong \textit{et al.}~\cite{xiong20235g} demonstrated that passive sniffing of the ABS–CN wireless backhaul in a 6G UAV testbed can expose app usage: by capturing GTP-encapsulated flows and applying CNN classifiers, they achieved $>$97\% accuracy in identifying applications despite encryption.
More recently, Li \textit{et al.}~\cite{li2026robust} introduced \emph{Crosshair}, a passive application-fingerprinting attack that observes the uplink rather than the downlink. The eavesdropper energy-detects PUSCH/PUCCH channel occupancy from raw IQ samples without decoding the RNTI-scrambled DCI. Because it targets the UE's transmissions, the attack does not depend on the gNodeB's downlink beam pattern. Its reception instead depends on the UE's transmit antenna pattern, its uplink power control, and its distance from the eavesdropper. In the configuration evaluated by the authors, this allowed the attack to avoid the two physical-layer barriers discussed in Section V-A: downlink beam selectivity and RNTI-based blind decoding. This advantage should not be generalized to all deployments. NR also supports uplink beamforming, particularly in FR2 \cite{3gpp38213, 3gpp38214}, and low uplink transmit power reduces the range at which occupancy can be detected.

\subsubsection{About position tracking}\label{subsec:position-tracking}
A smaller but significant group of works focus on passive user localization, showing that even encrypted or control-plane signals can leak spatial information.

Ateniese \textit{et al.}~\cite{ateniese2015no} demonstrated that a passive adversary observing encrypted TLS traffic can infer a user’s approximate location by leveraging flow‐level size and timing patterns; their approach shows that metadata alone may suffice for coarse geographic positioning. Complementing this approach, Yang \textit{et al}~\cite{yang2016passive} presented a passive localization approach for WiFi clients using RSSI (Received Signal Strength Indicator\footnote{RSSI represents the measured power level of a received radio signal, typically expressed in decibel-milliwatts (dBm). It is used by wireless systems (e.g., WiFi, LTE, 5G) to estimate link quality, signal coverage, and proximity to the transmitter.})-based fingerprinting combined with signal attenuation modeling and optimized fingerprint matching. They demonstrate via simulations and field experiments that it is possible to localize clients from intercepted signal strength measurements (i.e. without client cooperation), achieving both coarse “symbolic” localization and finer-grained positioning under favorable conditions.

Other studies moved deeper into the cellular stack. Kohls \textit{et al.}~\cite{kohls2019lost} performed passive layer-2 fingerprinting (MAC / RLC / PDCP) on LTE traffic to both identify accessed websites and map radio identities to users (identity mapping). Their work shows that link-layer metadata (e.g. packet size, sequence numbers, scheduling via RNTI) leaks discernible patterns even when higher-layer encryption is in place. In a similar line, Rupprecht \textit{et al.}~\cite{rupprecht2019breaking} uncovered LTE link-layer vulnerabilities by passively extracting layer-2 identifiers (e.g. RNTI/TMSI\footnote{Radio Network Temporary Identifier and Temporary Mobile Subscriber Identity}) and metadata-based signals (e.g. scheduling behavior), enabling identity mapping and website fingerprinting that compromise user anonymity under encryption.

Building upon these findings, Kotuliak \textit{et al}~\cite{kotuliak2022ltrack} introduced LTRACK, a fully passive LTE tracking attack that combines Timing Advance (TA) and Time of Arrival (ToA) measurements from uplink/downlink sniffer observations to localize user devices with sub-cell precision (e.g. $<$ 6 m error in line-of-sight). In the 5G context, Ludant \textit{et al.}~\cite{ludant20235g} developed 5GSniffer, the first open-source 5G control channel sniffer capable of decoding PDCCH/DCI messages in live NR networks, and demonstrated attacks that leverage these leaks, such as linking RNTIs to users and revealing their presence within a cell, while highlighting that control-channel vulnerabilities can still be exploited in modern 5G deployments.
Extending this line of work to the MAC and physical layers more broadly, Ludant \textit{et al.}~\cite{ludant2024unprotected} conducted a systematic study of unprotected 4G/5G control procedures, showing that beam-reporting and other low-layer signaling, unencrypted and unauthenticated even for connected, authenticated users, leaks enough spatial information to passively localize users to within 20 m with 96\% accuracy and to track their movement paths with 90\% reliability, all without decrypting any user-plane traffic. This result demonstrates that beamforming configuration itself, not just control-channel scheduling metadata, constitutes an exploitable passive side-channel for tracking.

\subsection{Data classification} \label{subsec:data-classification}
\begin{figure}[t]
    \centering
    \includegraphics[width=\linewidth]{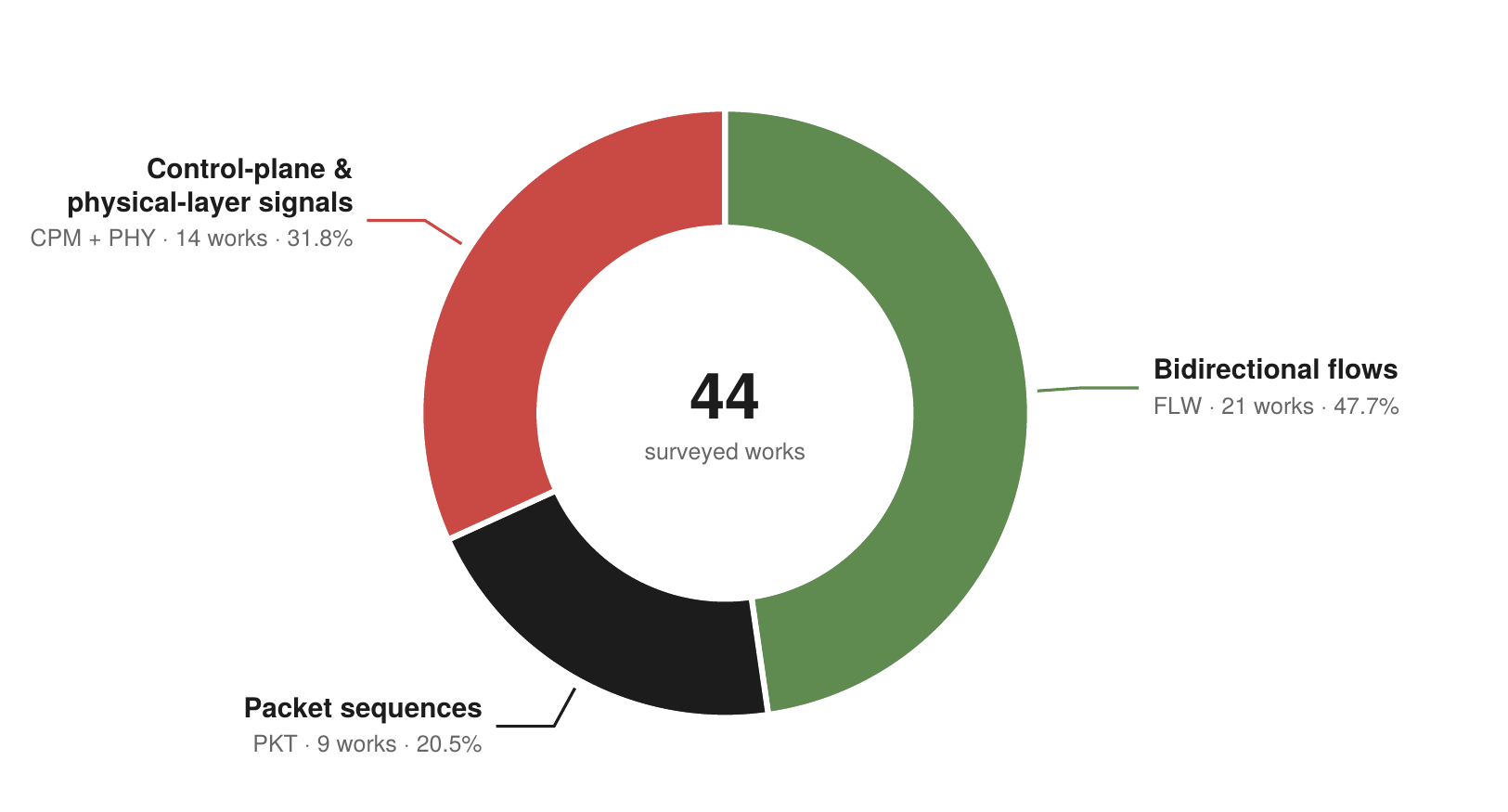}
    \caption{Proportion of the number of works per input data type.}
    \label{fig:classification-chart}
\end{figure}

We noticed that the presented works primarily splits into two main categories based on the nature of the input used by the attack: works based on simple packet sequences, and works based on structured flow-level data. But with the emergence of 5G and beyond, a third distinct family is increasingly present; relying on control-plane or physical-layer features (see Fig. \ref{fig:classification-chart} for a statistic).
%\vspace{-0.75\topsep}

\subsubsection{Packet sequences as input} This represents a small minority of all works that assumes as input the simplest and easiest to acquire side-channel information: a \textit{sequence of packet} lengths and times. That is a sequence of tuples of the form: $p_i = \langle t_i, s_i, d_i\rangle$ where the $i$-th captured packet is only represented by its timestamp $t_i$ with microsecond precision, a size in bytes $s_i$ and a binary direction $d_i$ (that is either uplink or downlink). A packet \textit{burst} is often defined as a subset of a packet sequence $[p_i, p_j]$ so that all packets are transmitted within some predefined time threshold, that is $t_j-t_i \leq \theta$. Bursts may alternatively be defined by characterizing their unusual high bitrate in comparison to average traffic, e.g. identifying $[p_i, p_j]$ as part of a downlink burst if
$$\frac{1}{t_j-t_i} \left( \sum_{\substack{i \leq k \leq j\\d_k = \text{downlink}}} s_k \right)> \theta,$$
where $\theta$ can be constant or made dependent on average transfer rate.

Several works from Table \ref{tab:pna-summary} rely exclusively on such packet-level metadata, without reconstructing sessions or requiring transport or network-layer headers. Early examples include Barbera \textit{et al.}~\cite{barbera2013signals} and Stöber~\cite{stober2013you}, who exploited Wi-Fi probe request frames and LTE/3G background packet timings, respectively, to infer user presence or device identity from raw packet traces. Similarly, Wang \textit{et al.}~\cite{wang2015know} and Ruffing \textit{et al.}~\cite{ruffing2016smartphone} used packet sizes and interarrival times to identify mobile applications or operating systems directly from encrypted wireless traffic.

A number of later studies further refined this packet-level approach by focusing on burst structures. Reed \textit{et al.}~\cite{reed2016leaky} analyzed encrypted HTTP adaptive streaming (DASH) traffic, showing that pieces of information such as packet sizes can uniquely identify video streaming content. Björklund's subsequent large-scale extension~\cite{usenixEndangeredPrivacy} confirmed that such burst fingerprints remain effective even when traffic is tunneled through VPNs or over different network technologies. 

Other packet-sequence attacks focus on user or device behavior inference. Saltaformaggio \textit{et al.}~\cite{saltaformaggio2016eavesdropping} detected in-app user activities by observing encrypted traffic patterns, while Acar \textit{et al.}~\cite{acar2020peek} applied similar reasoning to smart-home environments, where device state changes generate distinctive packet bursts across Wi-Fi, ZigBee, and BLE. Finally, Marañón \textit{et al.}~\cite{10823417} demonstrated that encrypted 5G traffic retains similar burst-level side channels, where only packet timing and size sequences are sufficient for mobile application classification.

\subsubsection{Bi-directional flows as input}
These represent the majority of the literature and in many cases, larger-scale results~\cite{rezaei2019large} have been obtained. 
Those works take as input network flows, which is requiring packet capture with at least clear IP headers. This information is for example not available to a simple eavesdropper attacker on a wireless channel encrypted at the MAC layer but requiring network infiltration or network administration rights.
Some of these works hence depart from the attack-perspective and rather take an administrator point of view intending to replace \textit{DPI} and infer traffic from its own network beyond application-layer or transport-layer encryption.
Most use \textit{bidirectional flows}, where traffic in both direction is aggregated in the same stream (i.e., source IP/port and destination IP/port are interchangeable).

A first family of works focuses on OS, browser, or device fingerprinting using network flow metadata. For instance, Chen \textit{et al.}~\cite{chen2014fingerprinting} showed that operating systems could be identified from ISP-level TCP/IP flow headers, while Muehlstein \textit{et al.}~\cite{muehlstein2017analyzing} and Rimmer \textit{et al.}~\cite{rimmer2018automated} applied machine learning to HTTPS and Tor flows to infer client OS, browser, and visited websites from encrypted traffic. These approaches rely on aggregated session-level features such as packet counts, directions, and timing distributions, extracted from bidirectional flows rather than from isolated packet sequences.

A large group of works targets mobile application identification from encrypted flow features. Taylor \textit{et al.}~\cite{taylor2016appscanner} identified Android apps from bidirectional TLS flow characteristics, later refined by Aceto \textit{et al.}~\cite{aceto2019mobile,aceto2020toward} through DL architectures operating on full Wi-Fi or LTE flows. Similarly, Rezaei \textit{et al.}~\cite{rezaei2019large} and D’Angelo \textit{et al.}~\cite{d2021network} leveraged CNN–LSTM and autoencoder models on per-flow features (packet size, direction, and inter-arrival time) to classify encrypted traffic at scale. Follow-up works such as Wang \textit{et al.}~\cite{wang2020automatic}, Gijón \textit{et al.}~\cite{gijon2020encrypted}, Montieri \textit{et al.}~\cite{montieri2021packet}, and Zhao \textit{et al.}~\cite{zhao2021optimized} explored both supervised and unsupervised learning on bidirectional flows to predict app activities, cluster traffic, or recognize services from transport-layer metadata.

Several other works examined encrypted traffic under realistic network settings and routing layers. Petagna \textit{et al.}~\cite{petagna2019peel} analyzed Tor traffic by reconstructing TCP flows to deanonymize Android apps, while Shapira \textit{et al.}~\cite{shapira2019flowpic} converted bidirectional flows into two-dimensional temporal–size histograms for CNN classification. Ateniese \textit{et al.}~\cite{ateniese2015no} demonstrated that coarse user location inference is possible from TLS-encrypted flow metadata alone, and Cheng \textit{et al.}~\cite{cheng2023watching} reconstructed VoLTE/VoNR call flows to recover session-level metadata in LTE/5G settings. Using the same relay-based interception principle, Wang \textit{et al.}~\cite{wang2026app} obtained genuine plaintext bidirectional flow metadata by decrypting PDCP traffic in transit through a MITM relay, rather than inferring flow characteristics from an encrypted side-channel. Finally, Xiong \textit{et al.}~\cite{xiong20235g} analyzed GTP/IP bidirectional flows in 6G UAV backhauls for application recognition.

Finally, the works of Dubin \textit{et al.}~\cite{dubin2017know}, Reed and Kranch~\cite{reed2017identifying}, and Björklund \textit{et al.}~\cite{10060390}, which focused on video traffic identification, grouped packets belonging to a single streaming session into a flow using transport identifiers (such as IP/port tuples and TLS/HTTP session metadata). From these grouped packets, the authors extract ordered burst sequences capturing burst timing and burst size in byte.

\subsubsection{Control-plane and physical-layer inputs} While most classical approaches rely on packet or flow-level metadata, a distinct class of recent works has emerged that bypasses traditional network-layer inputs entirely and uses non user-plane data. These attacks exploit control-plane messages or physical-layer features (such as radio signal structure, scheduling metadata, or uplink feedback) to infer user activity or presence. This is especially relevant in 5G and beyond, where PDCP encryption, beamforming, and decentralized architectures make extracting flows or even packets increasingly difficult.

\paragraph{Control-plane inputs} These studies operate at the signaling or protocol-control level and use decoded messages exchanged between the user equipment and the network (e.g., RRC, NAS, DCI, or PDCCH). They generally require software-defined radios or access to decoded traces but provide insights unavailable at the user plane. Early LTE-focused efforts such as Kohls \textit{et al.}~\cite{kohls2019lost} and Rupprecht \textit{et al.}~\cite{rupprecht2019breaking} revealed that even encrypted LTE traffic leaks identity and browsing information through link-layer scheduling and control identifiers. 
Later, Meneghello \textit{et al.}~\cite{meneghello2020smartphone} and Trinh \textit{et al.}~\cite{trinh2020mobile} analyzed the LTE downlink control channel (PDCCH) to fingerprint devices and identify applications directly from decoded DCI messages. 
With the advent of 5G, this line of work intensified: Budykho \textit{et al.}~\cite{budykho2023fine} demonstrated user linkability via RRC signaling analysis, while Wan \textit{et al.}~\cite{wan2024nr} built a passive 5G NR decoder that extracts telemetry and scheduling metadata from PDCCH/DCI messages. 
Similarly, Wani \textit{et al.}~\cite{wani2024security} exploited 5G NSA attach procedures to expose persistent identifiers from RRC/NAS exchanges, and Ludant \textit{et al.}~\cite{ludant20235g} showed that user presence can be detected by monitoring decoded NR control messages. Extending control-plane analysis further down the stack, Ludant \textit{et al.}~\cite{ludant2024unprotected} showed that lower-layer MAC and PHY control procedures, such as MAC Control Elements and beam-reporting messages, which fall outside the RRC/NAS scope of the works above, remain unencrypted and unauthenticated, and can be passively exploited to localize and track users.

\paragraph{Physical-layer inputs}
At an even lower layer, several attacks rely directly on radio signal features or measurements rather than protocol messages. 
Yang \textit{et al.}~\cite{yang2016passive} first demonstrated that Wi-Fi clients can be localized using RSSI fingerprints without any packet or flow information. 
Kotuliak \textit{et al.}~\cite{kotuliak2022ltrack} extended this idea to LTE by inferring user position from physical-layer timing metrics such as Timing Advance and Time-of-Arrival. 
More recent 5G studies leverage wideband SDR capture and physical feedback channels: Jawne \textit{et al.}~\cite{jawne2025ai} used deep learning on RF spectrograms to fingerprint individual 5G devices, while Zhang \textit{et al.}~\cite{zhang2025passive} extracted HARQ ACK/NACK sequences from the 5G PUCCH to infer application-level activity. Taking a different physical-layer vantage point, Li \textit{et al.}~\cite{li2026robust} observed the uplink rather than the downlink, showing that PUSCH/PUCCH channel occupancy recovered via raw IQ energy detection reliably reflects IP packet length under standard link adaptation, without requiring any message decoding.

\section{Feasibility Analysis}
\label{sec:feasbility}
This section presents one of the core contributions of our paper: a feasibility analysis of PNAs schemes in 5G/B5G environments.

\begin{table*}[t]
\begin{center}
\caption{\small Tool availability and cost across LTE, 5G, and B5G.}
\label{tab:tools}
\footnotesize
\begin{tabular}{p{1.5cm} p{2.5cm} p{4cm} p{1.5cm} p{3cm} p{2cm} p{1.5cm}}

\hline 

\textbf{Capture task} & \textbf{Frequency coverage} & \textbf{Instantaneous BW (min. sample rate)} & \textbf{RF chains} & \textbf{Example hardware} & \textbf{Open-source tooling} & \textbf{Indicative price} \\

\hline
Wi-Fi monitor mode & 2.4/5 GHz ISM & 20--80 MHz (handled by NIC) & NIC-dependent & ALFA AWUS036ACH \cite{alfa_awus036ach} & Wireshark, Kismet & \$65 \\

LTE PDCCH/DCI sniffing & Sub-6 GHz & $\leq$20 MHz (30.72 MS/s) & 1--2 & bladeRF 2.0 micro xA4 \cite{nuand_xa4}; USRP B210 \cite{ettus_b210} & LTESniffer, OWL & \$540; \$2,387 \\

NR SSB / CORESET 0 / narrow BWP & FR1 (sub-6 GHz) & 7.2--34.56 MHz ($\leq$61.44 MS/s) & 1 & USRP B210 \cite{ettus_b210} & 5GSniffer, NR-Scope & \$2,387 \\

NR full FR1 carrier IQ & FR1 (sub-6 GHz) & 100 MHz (122.88 MS/s) & 1--2 & USRP X310 + UBX-160 \cite{ettus_x310,ettus_ubx160}; USRP N310 \cite{ettus_n310} & Custom only & \$14,107; \$20,826 \\

NR FR2 & FR2-1: 24.25--52.6 GHz; FR2-2: 52.6--71 GHz \cite{3gpp381012} & Up to 400 MHz (491.52 MS/s) & $\geq$1 + mmWave front-end & USRP X410 \cite{ettus_x410} + mmWave converter & None public & \$33,020 + converter (quote) \\

B5G/6G upper mid-band & 7.125--8.4 GHz \cite{sampath2024imt2030} & Not yet standardized & Array-based & USRP X420 \cite{ettus_x420} & None public & \$52,920 \\

B5G/6G sub-THz & $>$100 GHz & Not yet standardized & Array-based & Research prototypes & None public & Quote-based \\

\hline

\end{tabular}
\end{center}
\end{table*}

\subsection{Feasibility factors} \label{subsec:factors}
The objective of the survey is to produce a feasibility analysis of the difficulty of reproducing known PNAs in 5G/B5G networks. 
While many fingerprinting and inference attacks have been successfully demonstrated in WiFi and LTE contexts, the jump to 5G and B5G introduces new challenges and opportunities that impact attacker capabilities.
All the following aspects of practicality for an attacker to perform the attack are explored and analyzed (see Table \ref{tab:tools} for a summary.):
\begin{itemize}[itemsep=0.5pt,topsep=0.5pt]
    \item Is the needed hardware easily available on the market? Since commercial LTE and even more 5G devices do not offer network monitoring functions, a custom-based setting (antenna within appropriate frequency range and some software-defined radio code) is likely required without access to the very few closed-source professional equipment. 
    \item 5G radio communications are much more spatially selective than in previous systems. How close does an eavesdropper need to be to capture enough data? 
    \item 5G antennas are deployed closer to the users. Can this facilitate privacy leaks following the popular belief, or, on the contrary, make the reproduction of known attacks an even more challenging task?
\end{itemize}

In earlier work, many attacks targeted WiFi-connected devices, largely due to ease of access and high bandwidth. WiFi traffic can be captured using commodity devices. In contrast, cellular traffic, particularly in 5G, poses several physical, protocol-layer, and resource constraints that affect attack reproducibility and scale.

\subsubsection{Physical Layer Constraints: The Beamforming Barrier}\label{subsec:phy-constraints}
\begin{figure*}[t]
    \centering
    \includegraphics[width=\linewidth]{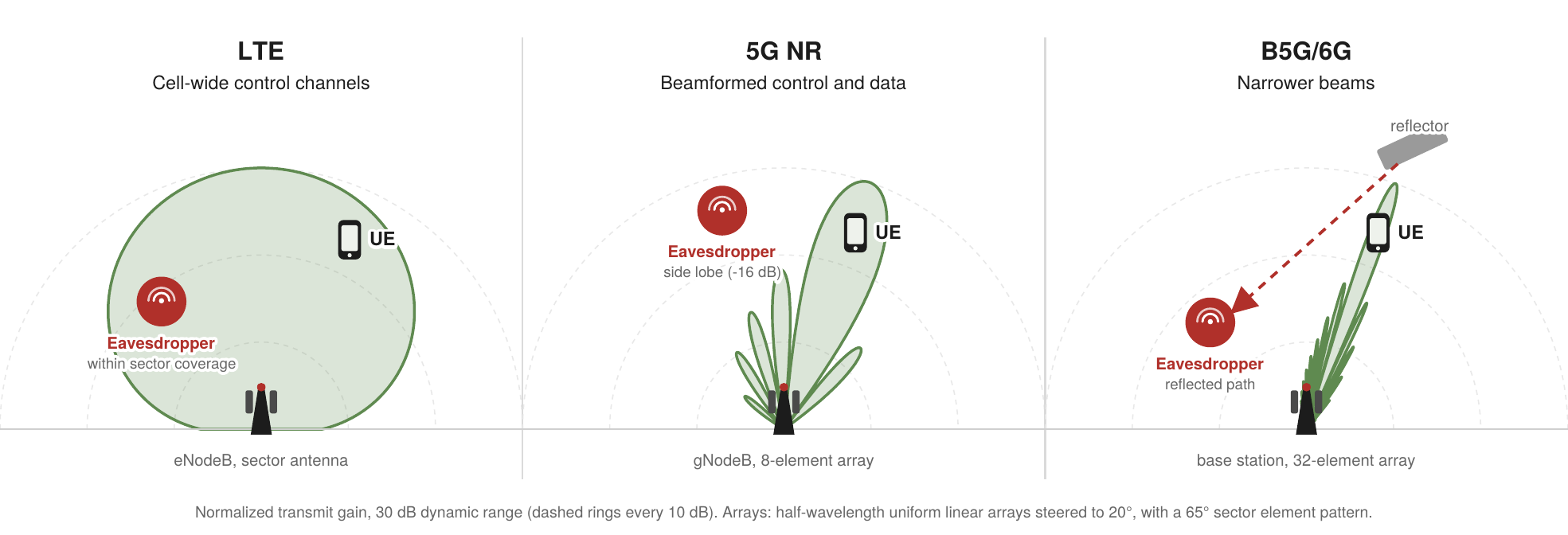}
    \caption{Illustrative downlink transmission patterns. LTE control channels are typically transmitted cell- or sector-wide, while NR integrates beamforming into initial access and control-channel transmission. Actual angular selectivity depends on carrier frequency, array configuration, geometry, and channel conditions.}
    \label{fig:mimo}
\end{figure*}

Unlike Wi-Fi access points, which typically transmit with broad antenna patterns, 5G NR gNodeBs, particularly in FR2, use antenna arrays to concentrate transmitted energy toward the scheduled UE. Whether a passive eavesdropper (Evsdr) can decode a given transmission therefore depends on the transmit array pattern, the propagation environment, its distance to the gNodeB, and its own receive gain (see Figure \ref{fig:mimo} for an overview).

In a line-of-sight setting, the power received by Evsdr can be modeled by extending the Friis transmission equation~\cite{friis1946note}:
$$P_{E}= P_{T} + G_{T}(\theta_{E}) + G_{R,E} - PL_{E}
$$
Where:
\begin{itemize}
    \item $P_{T}$ is the gNodeB transmit power.
    \item $G_{T}(\theta_{E})$ the transmit array gain toward Evsdr.
    \item $G_{R,E}$ Evsdr's receive gain.
    \item $PL_{E}$ the path loss to Evsdr (all in dB/dBm).
\end{itemize}
In multipath channels, the term $G_{T}(\theta_{E}) - PL_{E}$ is replaced by the effective beamformed channel gain $G_{eff,E} = 10 \log_{10}|h^H_Ew|^2$, where $h_E$ is the channel vector from the array to Evsdr and $w$ the precoding vector applied for the target UE.
\medskip
Two important notions here have to be mentioned.

\paragraph{Main-lobe, Side-lobe, and Multipath Reception} An eavesdropper at a different angle from the target (i.e $\theta_{Evsdr} \neq \theta_{UE}$) is not thereby outside the main lobe. A beam has a finite width, determined by array size, carrier frequency, and steering angle, so an eavesdropper at a nearby angle may still receive close-to-peak gain. Further off-axis, Evsdr relies on side-lobe gain. Its level depends on the array size, element pattern, amplitude taper (e.g., uniform, Chebyshev, or Taylor weighting), and steering angle. First side-lobe levels around 13 dB below the peak are typical of uniformly weighted arrays, and tapering can lower them further at the cost of a wider main lobe. The side-lobe level is therefore configuration-dependent rather than a fixed property of 5G arrays. In multipath channels, angular separation alone is insufficient, since a reflector can redirect main-lobe energy toward an eavesdropper located far from the target direction; this is captured by $G_{eff,E}$ rather than by $G_{T}(\theta_{E})$. Consequently, no single compensation figure applies. Whether Evsdr can offset a lower transmit gain through proximity or receive gain must be assessed for the specific array, geometry, and channel.
\paragraph{Decoding Threshold} To exploit a transmission, Evsdr must decode it, for example the DCI carried on the PDCCH. This requires its received SNR to exceed the decoding threshold of the selected transmission format (modulation and coding scheme, or PDCCH aggregation level) for its receiver. Under an additive, noise-limited model, for the same transmitted resource, the SNR difference between Evsdr ($E$) and the target UE ($U$) is
$$
\begin{aligned}
\mathrm{SNR}_E-\mathrm{SNR}_U ={}& [G_T(\theta_E)-G_T(\theta_U)] + (G_{R,E}-G_{R,U}) \\
&+ (PL_U-PL_E) + (N_U-N_E),
\end{aligned}
$$
where $G_T$ and $G_R$ are transmit and receive gains, $PL$ is path loss, and $N$ is noise power. SNRs, gains, and path losses are in dB, and noise powers in dBm.

Each term can favor either party. A shorter distance to the gNodeB, a higher-gain receive antenna, or a lower receiver noise floor can offset a lower transmit gain toward Evsdr.

The simplified relation $\mathrm{SNR}_E \approx \mathrm{SNR}_{Target} - (G_{main} - G_{side})$ holds only under all of the following conditions:

path loss, receive gain, and noise power are equal for both receivers;
the UE lies at the beam peak;
Evsdr lies in a side lobe;
the UE operates exactly at $\mathrm{SNR}_{Target}$.

Moreover, the gNodeB does not in general reduce transmit power to just meet a fixed target SNR. Link adaptation selects the transmission format and the power allocation per resource, so the margin available to Evsdr depends on both. When $\mathrm{SNR}_E$ falls below the decoding threshold of the selected format, the transmission cannot be decoded reliably regardless of computational power. Favorable geometry, multipath, or receive gain can, however, bring it above that threshold, so decodability is a configuration-dependent question rather than a physical impossibility.

\subsubsection{Protocol Layer Barriers: The "Blind Decoding" Complexity}
Even when Evsdr captures the physical signal, 5G NR imposes a protocol-layer barrier to interpreting it. Unlike Wi-Fi, where MAC headers are transmitted in the clear, the NR downlink control channel must be blindly decoded, and the identity needed to bind a transmission to a user is not exposed directly. 

\paragraph{The RNTI Scrambling} The 16-bit CRC of a DCI message is scrambled by an XOR with the C-RNTI. A common misconception is to think that a sniffer tests 65,536 candidate RNTIs. In contrast, after decoding a PDCCH candidate it computes the CRC over the payload and XORs it with the received parity bits, which yields the candidate RNTI directly in a single operation. This is precisely how open LTE monitors (OWL \cite{bui2016owl}, FALCON \cite{falkenberg2019falcon}) and the NR tools (NR-Scope \cite{wan2024nr}, 5GSniffer \cite{ludant20235g}) run in real time on general-purpose CPUs. The cost per slot is therefore one channel decode per blind-decoding candidate, not per RNTI.

\paragraph{The Cost} The number of monitored PDCCH candidates per slot per serving cell is capped by 3GPP (TS 38.213, \cite{3gpp38213}): at most 44 for $\mu=0$ (15 kHz SCS), $36$ for $\mu=1$ (30 kHz), $22$ for $\mu=2$, and $20$ for $\mu=3$, with a parallel limit on non-overlapping CCEs (56/56/48/32). A sniffer performs on the order of a few tens of polar decodes per slot across aggregation levels and candidate DCI sizes; which is within reach of a modern GPP, as the above tools demonstrate.

\paragraph{The RNTI Validation} Because the CRC-XOR yields some 16-bit value for every candidate, any decode ``passes.'' Distinguishing a genuine C-RNTI from a false positive is the hard part: sniffers must bootstrap RNTIs from the random-access procedure (RA-RNTI $\rightarrow$ TC-RNTI $\rightarrow$ C-RNTI at Msg4) or validate them through cross-slot consistency (repeated scheduling, plausible DCI formats and aggregation levels). This bootstrapping, not raw compute, is what makes turnkey passive capture difficult and what required manual calibration in \cite{wan2024nr, ludant20235g}.

\paragraph{NR-specific Hardening} In a UE-specific search space, NR can initialise both the PDCCH data scrambling (TS 38.211 \cite{3gpp38211}) and, optionally, the PDCCH DMRS with a configured scrambling identity and the C-RNTI, rather than with the physical cell ID as in common search spaces and CORESET 0. Where this is configured, descrambling a UE-specific PDCCH requires the C-RNTI before decoding, so the CRC-XOR recovery above no longer applies and the attacker is confined to common search spaces and CORESET 0 unless the target’s C-RNTI is already known. This is a genuine step up from LTE, where PDCCH scrambling is cell-ID based throughout. It also means control-plane sniffing is easiest exactly where transmission is cell- or sector-wide (SSB, CORESET 0, common PDCCH), which partly offsets the beamforming barrier of Section \ref{subsec:phy-constraints} for those channels.

\paragraph{Packet-level visibility below PDCP} PDCP ciphering encrypts the PDCP payload but not the control information that schedules and frames it. Two routes to packet-level side-channels therefore survive in 5G:
\begin{enumerate}
    \item \textbf{From DCI only.} The transport-block size and timing carried in each DCI expose the per-slot downlink/uplink volume and cadence. Trinh \textit{et al.} \cite{trinh2020mobile} exploit exactly this for LTE app/service classification, and Bae \textit{et al.} \cite{bae2022watching} use DCI-derived TBS sequences for practical passive video identification in LTE (a close analogue of the burst- and size-based attacks we classify in this survey). Equivalent PDCCH/DCI decoding has since been demonstrated on live 5G NR \cite{wan2024nr, ludant20235g}, so this route is available in principle on 5G.
    \item \textbf{From the decoded PDSCH.} Once the C-RNTI and DCI are known and the PDSCH is decoded, the MAC subheaders and RLC headers, including SDU lengths, are not PDCP-ciphered. This is the mechanism underlying the LTE layer-2 attacks of Kohls \textit{et al.} \cite{kohls2019lost} and Rupprecht \textit{et al.} \cite{rupprecht2019breaking}.
\end{enumerate}

\subsubsection{Resource Barriers: Hardware Cost and Software Maturity} \label{subsec:hardware-cost}
Beyond physics and algorithms, practical feasibility is severely limited by the availability of Commercial Off-The-Shelf (COTS) tools. We show here a comparison of available tools for different network types (see Table \ref{tab:tools} for a summary).

\paragraph{Acquisition Bandwidth and Recording Throughput} NR carriers span up to 100 MHz in FR1 and up to 400 MHz in FR2 \cite{3gpp381012}. The capture capability required depends on the observation task (see Table \ref{tab:tools})
\begin{itemize}
    \item \textbf{SSB or CORESET 0:} roughly 7–35 MHz at 30 kHz subcarrier spacing, within reach of mid-range SDRs.
    \item \textbf{Full 100 MHz FR1 carrier:} requires at least 122.88 MS/s (the standard rate for a 4096-point FFT at 30 kHz subcarrier spacing). This exceeds the B210's capabilities \cite{ettusb2x0} and calls for platforms such as the USRP X310 or N310 with a 10 Gigabit Ethernet link.
\end{itemize}
\textbf{Recording throughput.} At 122.88 MS/s, one receive channel produces 491.52 MB/s when stored as 16-bit complex integers (sc16, 4 bytes per sample) and 983.04 MB/s as 32-bit complex floats (fc32, 8 bytes per sample). Each additional RF chain multiplies these figures. This recording rate is distinct from interface throughput: an sc16 stream occupies about 3.9 Gb/s of a 10 GbE link per channel, while the storage subsystem must sustain the chosen stored format, which limits the duration of offline captures.

\paragraph{Lack of "Promiscuous" Software Stacks} As noted in~\cite{kohls2019lost}, in the LTE/4G domain, robust open-source stacks such as srsLTE integrated with passive sniffers like Airscope, OWL~\cite{rupprecht2019breaking}, and LTESniffer enable researchers to capture real cell traffic with modest setups. These tools run on general-purpose SDR platforms such as the USRP B200/B210 or bladeRF, since a 20 MHz LTE carrier fits within their instantaneous bandwidth (see Table \ref{tab:tools} for dated prices).
In the 5G context, although sub-6 GHz SDRs, including multi-antenna units (e.g., bladeRF 2.0 micro, USRP B2xx/B210), are OTS; general-purpose mmWave front-ends and phase-coherent arrays remain specialized and scarce. 

Open-source tools like srsRAN or OpenAirInterface are designed as endpoints (UE or gNodeB). They operate state machines that expect a specific assigned RNTI. Modifying these tools for "promiscuous mode" sniffing is non-trivial because they lack the architecture for parallelized, massive blind decoding across the entire RNTI space. This explains the scarcity of "plug-and-play" 5G sniffers compared to the Wi-Fi ecosystem.

Capturing PDCCH control-plane messages remains however possible, but only via custom-developed tools or patched forks of LTE-era decoders. For example, \cite{ludant20235g} successfully decoded DCI messages to track user presence in a live 5G NSA cell, while \cite{wan2024nr} extracted RRC scheduling telemetry from the 5G downlink via specialized PDCCH sniffers. These works required manually configured SDRs and tuning of decoder parameters, often relying on adapted versions of OWL or private tools layered atop srsRAN.
Other recent works such as \cite{jawne2025ai} bypass the network stack entirely and capture RF signals at the physical layer, using spectrogram-based fingerprinting to distinguish between 5G user devices. This approach requires wideband RF sampling hardware and DL-capable GPUs for inference, increasing cost and system complexity. Similarly, \cite{wani2024security} demonstrated passive monitoring of real-world 5G NSA deployments to uncover IMSI leakage and tracking vulnerabilities, using adapted SDR toolchains.

\paragraph{The move to B5G/6G} The situation in B5G and 6G is even more constrained. Most experimentation is limited to research testbeds using non-public platforms or hardware prototypes, with no general-purpose open-source passive monitoring tools currently available.

\subsection{Demonstrability of the surveyed works on 5G}\label{subsec:dempónstrability}
We start here by analyzing whether the surveyed PNAs in Section \ref{subsec:data} can be reproducible on 5G networks. This helps us situate where the current literature is headed. We consider in our assessment a passive and external attacker. We classified the works into 4 levels of feasibility (see Table \ref{tab:pna-feasibility} for a summary). To make this classification reproducible, we apply the following decision procedure to each surveyed work, in order:
\begin{enumerate}
    \item \textbf{Q1: Data accessibility.} Does the attack require bidirectional flow-level visibility (clear IP/TCP headers) or core-network-side data (e.g., a GTP or ISP vantage point)? If yes, we distinguish two cases based on how this visibility was obtained. If the flow-level data was recovered by passively observing an already end-to-end-ciphered channel, the work is assigned \textbf{Class 1 (Not reproducible / Not applicable)}, since this data is categorically hidden behind PDCP ciphering and GTP encapsulation in 5G, with no known passive bypass. If instead the flow-level data was obtained through active relay-based interception, i.e., a mobile-relay or Man-in-the-Middle setup that decrypts PDCP-layer traffic in transit before forwarding it to the legitimate network (e.g., Cheng \textit{et al.} (2023)~\cite{cheng2023watching}, Wang \textit{et al.} (2026)~\cite{wang2026app}), the passive-adversary assumption underlying the Class 1 case does not hold, since the relay operator obtains genuine plaintext by construction rather than inferring it from a ciphered side-channel; such works instead proceed to Q3, with their input treated as CPM-equivalent. If no (the attack does not require flow-level or core-network-side data at all), proceed to Q2.
    \item \textbf{Q2: Input layer and layer of visibility.} %Is the required input a packet sequence (user-plane packet timing and size, requiring PDCP-frame demodulation and reconstruction to extract), or a control-plane message / physical-layer signal (e.g., DCI, RRC, RF features, observable without decrypting user-plane content)? Packet-sequence inputs are capped at \textbf{Class 2 (Reproducible with extensive efforts)}, regardless of whether the specific work has been tested on 5G, since extracting packet-level metadata from ciphered PDCP frames always requires the same extensive SDR-based demodulation and reconstruction effort. Control-plane and physical-layer inputs proceed to Q3.
    If the attack does not require flow- or core-network-side data, we ask how its packet-level input is obtained, because PDCP ciphering conceals only the PDCP payload (the IP packet), not the scheduling metadata or the lower-layer headers that frame it. We distinguish two cases:
    \begin{enumerate}
        \item \textbf{Metadata requiring user-plane frame reconstruction.} If the attack genuinely needs the reassembled user-plane byte stream (e.g. exact PDCP SDU boundaries recovered by decoding the PDSCH), it is capped at Class 2, since that reconstruction demands full-carrier capture, RNTI/DCI recovery, PDSCH decoding, and handling of the PDSCH barriers discussed below.
        \item \textbf{Metadata derivable without decoding user-plane content.} Per-slot transport-block sizes, scheduling times, and HARQ activity can be read from the DCI alone, yielding a sequence of (size, time, direction) tuples that is a faithful proxy for the user-plane packet sequence without decrypting any PDCP frame. Such inputs are observable at the air interface and proceed to Q3 on the same footing as other control-/physical-layer inputs.
    \end{enumerate}
    \item \textbf{Q3: Demonstrated transfer and resource threshold.} Has the technique, or a direct equivalent, been experimentally validated on real 5G NR hardware or deployments? If not (only shown on LTE or Wi-Fi), the work is assigned Class 2. If demonstrated on 5G but requiring specialized or custom SDR tooling, manual calibration, or favorable positioning, the work is assigned \textbf{Class 3 (Reproducible with efforts)}. If demonstrated on 5G and reproducible with commodity hardware and turnkey open-source tools, with no proximity or beam-alignment constraint, the work is assigned \textbf{Class 4 (Easily reproducible)}.
\end{enumerate}

\subsubsection{Class 1: Not Reproducible / Not Applicable} This class includes attacks that cannot be reproduced on 5G at all, mainly because they depend on data that is fundamentally inaccessible under the 5G security model. For instance, some attacks require information from the physical layer of user equipment that is now fully encrypted and therefore out of reach for an external eavesdropper. Since no known tools, workarounds, or experimental methods exist to bypass these barriers, such attacks are considered theoretically impossible to reproduce in practical 5G scenarios.

A large subset of early works fall in this class because they rely on packet or flow-level visibility that 5G completely conceals behind PDCP encryption and GPRS Tunneling Protocol (GTP) encapsulation. These methods typically assume access to plaintext IP/TCP headers or reconstructed sessions, which an external 5G eavesdropper cannot observe.

\begin{itemize}
    \item Barbera \textit{et al.}~\cite{barbera2013signals} relies on Wi-Fi probe requests and management frames observable only in 802.11 environments, with no equivalent in 5G.
    \item Chen \textit{et al.}~\cite{chen2014fingerprinting}, Taylor \textit{et al.}~\cite{taylor2016appscanner}, Dubin \textit{et al.}~\cite{dubin2017know}, Muehlstein \textit{et al.}~\cite{muehlstein2017analyzing}, Rimmer \textit{et al.}~\cite{rimmer2018automated}, Aceto \textit{et al.}~\cite{aceto2019mobile, aceto2020toward}, Petagna \textit{et al.}~\cite{petagna2019peel}, Shapira \textit{et al.}~\cite{shapira2019flowpic}, Rezaei \textit{et al.}~\cite{rezaei2019large}, D’Angelo \textit{et al.}~\cite{d2021network}, Gijón \textit{et al.}~\cite{gijon2020encrypted}, Montieri \textit{et al.}~\cite{montieri2021packet}, and Zhao \textit{et al.}~\cite{zhao2021optimized} all require bidirectional IP/TCP flow information or transport-layer statistics. These inputs are encrypted and multiplexed in 5G, making passive capture and reconstruction impossible without operator access.
    \item Xiong \textit{et al.}~\cite{xiong20235g} and Ateniese \textit{et al.}~\cite{ateniese2015no} rely on GTP or ISP-level traffic features available only inside the core network, not observable over the air.
    \item Both burst-based methods of Reed \& Kranch~\cite{reed2017identifying} and Björklund \textit{et al.}~\cite{10060390} rely on clean and precise packet-level visibility (sizes, timings, ordering) obtainable in Wi-Fi or wired settings; this assumption is fundamentally impossible in LTE/5G due to protocol encapsulation and encrypted transport blocks.
\end{itemize}

\begin{table}[t]
    \centering
    \caption{Feasibility classification of passive network attacks}

    \begin{tabular}{c c c c}
        \hline
        \textbf{Authors (Year) [ref.]} & \textbf{Problem tackled} & \textbf{Input} & \textbf{on 5G} \\
        \hline
%        \small 

        Barbera \textit{et al.} (2013)~\cite{barbera2013signals} & Relation ID & PKT & \emptycircle \\

        Chen \textit{et al.} (2014)~\cite{chen2014fingerprinting} & OS ID & FLW & \emptycircle \\

        Ateniese \textit{et al.} (2015)~\cite{ateniese2015no} & Localization TR & FLW & \emptycircle \\

        Taylor \textit{et al.} (2016)~\cite{taylor2016appscanner} &  Application ID & FLW & \emptycircle \\ 

        Dubin \textit{et al.} (2017)~\cite{dubin2017know} & Video ID & FLW & \emptycircle \\
        
        Rimmer \textit{et al.} (2017)~\cite{rimmer2018automated} & Website ID & FLW & \emptycircle \\

        Muehlstein \textit{et al.} (2017)~\cite{muehlstein2017analyzing} & OS/browser/app. ID & FLW & \emptycircle \\

        Aceto \textit{et al.} (2019)~\cite{aceto2019mobile} & Traffic ID & FLW & \emptycircle \\

        Petagna \textit{et al.} (2019)~\cite{petagna2019peel} & Application ID & FLW & \emptycircle \\

        Shapira \textit{et al.} (2019)~\cite{shapira2019flowpic} & Traffic ID & FLW & \emptycircle \\
        
        Rezaei \textit{et al.} (2019)~\cite{rezaei2019large} & Application ID & FLW & \emptycircle \\

        D’Angelo \textit{et al.} (2019)~\cite{d2021network} & Traffic ID & FLW & \emptycircle \\

        Aceto \textit{et al.} (2020)~\cite{aceto2020toward} & Traffic ID & FLW & \emptycircle \\

        Wang \textit{et al.} (2020)~\cite{wang2020automatic} & Application ID & FLW &  \emptycircle \\

        Gijon \textit{et al.} (2020)~\cite{gijon2020encrypted} & Traffic ID & FLW & \emptycircle \\
        
        Montieri \textit{et al.} (2021)~\cite{montieri2021packet} & Pkt ID & FLW & \emptycircle \\

        Zhao \textit{et al.} (2021)~\cite{zhao2021optimized} & Traffic ID & FLW & \emptycircle \\

       Xiong \textit{et al.} (2023)~\cite{xiong20235g} & Activity ID & FLW & \emptycircle \\

        Reed \textit{et al.} (2017)~\cite{reed2017identifying} & Video ID & FLW & \emptycircle \\

        Björklund \textit{et al.} (2023)~\cite{10060390} & Video ID & FLW & \emptycircle \\

        \hline
        \hline
        
        Stöber (2013)~\cite{stober2013you} & User ID & PKT & \thirdpie \\

        Wang \textit{et al.} (2015)~\cite{wang2015know} & Application ID & PKT & \thirdpie \\
        
        Saltaformaggio \textit{et al.} (2016)~\cite{saltaformaggio2016eavesdropping} & Activity ID & PKT & \thirdpie \\

        Ruffing \textit{et al.} (2016)~\cite{ruffing2016smartphone} & OS ID & PKT & \thirdpie \\
         
        Reed \textit{et al.} (2016)~\cite{reed2016leaky} & Video ID & PKT & \thirdpie \\

        Acar \textit{et al.} (2020)~\cite{acar2020peek} & Activity ID & PKT & \thirdpie \\
       
       Marañón \textit{et al.} (2024)~\cite{10823417} & Application ID & PKT & \thirdpie \\

       Björklund \textit{et al.} (2025)~\cite{usenixEndangeredPrivacy} & Video ID & PKT & \thirdpie \\

       Yang \textit{et al.} (2016)~\cite{yang2016passive} & Passive TR & PHY & \thirdpie \\
    
       Kotuliak \textit{et al.} (2022)~\cite{kotuliak2022ltrack} & UE TR & PHY & \thirdpie \\

       Kohls \textit{et al.} (2019)~\cite{kohls2019lost} & User TR & CPM & \thirdpie \\
        
        Rupprecht \textit{et al.} (2019)~\cite{rupprecht2019breaking} & User ID & CPM & \thirdpie \\

        Meneghello \textit{et al.} (2020)~\cite{meneghello2020smartphone} & Smartphone ID & CPM & \thirdpie\\

        Trinh \textit{et al.} (2020)~\cite{trinh2020mobile} & App/Service ID & CPM & \thirdpie \\

       \hline
       \hline
        
       Ludant \textit{et al.} (2023)~\cite{ludant20235g} & Movement TR & CPM & \twothirdspie \\

       Budykho \textit{et al.} (2023)~\cite{budykho2023fine} & Entity ID & CPM & \twothirdspie \\

       Wan \textit{et al.} (2024)~\cite{wan2024nr} & Telemetry ID & CPM & \twothirdspie \\

       Wani \textit{et al.} (2024)~\cite{wani2024security} & IMSI ID & CPM & \twothirdspie \\

       Ludant \textit{et al.} (2024)~\cite{ludant2024unprotected} & User TR & CPM & \twothirdspie \\

       Cheng \textit{et al.} (2023)~\cite{cheng2023watching} & VoLTE ID & FLW* & \twothirdspie \\

       Wang \textit{et al.} (2026)~\cite{wang2026app} & Activity ID & FLW* & \twothirdspie\\

       Jawne \textit{et al.} (2025)~\cite{jawne2025ai} & Devices ID & PHY & \twothirdspie \\

       Zhang \textit{et al.} (2025)~\cite{zhang2025passive} & Application ID & PHY & \twothirdspie\\
        
       Li \textit{et al.} (2026)~\cite{li2026robust} & Application ID & PHY & \twothirdspie\\
       
        \hline
        \hline
        / & / & / & \fullcircle \\

        \hline
    \end{tabular}

    \smallskip
    \begin{footnotesize}
    \smallskip
    \textbf{Acronyms:}  ID = Identification, TR = Tracking, CPM = Control Plane Message, FLW = Flow, FLW* = Flow obtained through relay-based interception, PHY = Physical Layer Data, PKT = Packet Sequence.\\
    \textbf{Symbols:} \emptycircle {} = Not Reproducible, \thirdpie = Reproducible with extensive efforts, \twothirdspie {} = Reproducible with efforts, \fullcircle {} = Easily Reproducible.
    \end{footnotesize}
    \smallskip
    \label{tab:pna-feasibility}
\end{table}

\subsubsection{Class 2: Reproducible with extensive efforts} This class refers to attacks that, in principle, could be adapted to 5G but would require significant effort, technical expertise, and specialized equipment to achieve. Typically, these are techniques demonstrated on Wi-Fi or LTE that do not directly transfer to 5G without major modifications. They often require custom-built SDR pipelines, proprietary tools, or carefully controlled laboratory environments, which makes them impractical for most adversaries. While not impossible, the high cost and complexity place these attacks in a category of limited feasibility.

A large part of the surveyed works falls in this class. 
\begin{itemize}
    \item  Works such as Stöber~\cite{stober2013you}, Wang \textit{et al.}~\cite{wang2015know}, Ruffing \textit{et al.}~\cite{ruffing2016smartphone}, Saltaformaggio \textit{et al.}~\cite{saltaformaggio2016eavesdropping}, and Acar \textit{et al.}~\cite{acar2020peek} infer device or user behavior from packet timing and/or size sequences. While conceptually valid on 5G, reproducing these attacks requires complete SDR demodulation and extraction of user-plane packet events from PDCP frames, which is currently feasible only with custom signal-processing pipelines and precise synchronization.
    \item Yang \textit{et al.}~\cite{yang2016passive} exploits RSSI data from commodity Wi-Fi hardware; such per-packet PHY metrics are not available from 5G modems. However, 5G offers alternative signals that carry spatial information such as RSRP (Reference Signal Received Power, the measured power level of a 5G reference signal received by a device, indicating signal strength from a specific cell) and Beam indexes (identifiers for the directional beams used in 5G to guide radio signals between a base station and a device). Using specialized and advanced SDR gear can also help decoding synchronization sequences instead of traffic frames, which can enable high-resolution fingerprints. In other words, the attack as it is cannot be reproduced on 5G unless it is adapted to capturing spatial information from 5G-specific signals.
    \item The other works of Reed \textit{et al.}~\cite{reed2016leaky} and Björklund \textit{et al.}~\cite{usenixEndangeredPrivacy}; in contrast to the first two~\cite{reed2017identifying, 10060390}, depend on identifying characteristic download bursts in encrypted DASH or HTTPS traffic without requiring access to network, or having precise data visibility. The burst patterns used by these attacks (packet-burst timing and size features) still exist in 5G user-plane traffic but can only be accessed by isolating PDCP bursts through advanced SDR decoding and manual calibration, requiring significant effort.
    \item Kotuliak \textit{et al.}~\cite{kotuliak2022ltrack} exploits LTE Timing Advance and ToA for localization, and Marañón \textit{et al.}~\cite{10823417} proposes 5G app fingerprinting using burst-level side channels. Both require specialized SDR setups, multi-antenna synchronization, and precise timing recovery to be reproduced on 5G NR, making them feasible most likely only in carefully controlled environments.
    \item Meneghello \textit{et al.}~\cite{meneghello2020smartphone} and Trinh \textit{et al.}~\cite{trinh2020mobile} decode LTE control-plane information such as DCI and PDCCH messages to fingerprint user sessions. These works fall into this class as they have not yet been shown to work on 5G; however, it is important to note that equivalent PDCCH/DCI decoding has since been demonstrated on 5G NR testbeds using open-source SDR toolchains (e.g \cite{ludant20235g, wan2024nr}).
\end{itemize}

\subsubsection{Class 3: Reproducible with efforts} \label{subsec:class3}
This class contains attacks that have already been demonstrated on 5G, but only under specific conditions that demand non-trivial effort from the attacker. Examples include experiments on 5G testbeds or small scale live networks where techniques such as decoding control plane messages or extracting side channel features have been shown to work. While reproduction is possible, it usually requires mid-range SDR hardware, manual tuning, and favorable conditions such as sufficient received SINR or reduced noise. These attacks are feasible, but not straightforward, and their success depends heavily on the attacker’s expertise and setup.

Verily, only a few works fit in \emph{Class 3}:
\begin{itemize}
    \item Budykho \textit{et al.}~\cite{budykho2023fine} reveals user linkability through 5G RRC signaling, and Ludant \textit{et al.}~\cite{ludant20235g} detects 5G device presence from PDCCH scheduling activity. Both operate natively on 5G NR and can be reproduced with moderate effort using known decoding frameworks such as srsRAN or OAI.
    \item Cheng \textit{et al.}~\cite{cheng2023watching} demonstrates side-channel inference from VoNR session metadata, showing practical feasibility on operational 5G networks with limited lab instrumentation. Wang \textit{et al.}~\cite{wang2026app} apply this same relay methodology to demonstrate What-App?, a MITM-relay-based application-usage fingerprinting attack achieving 99.02\% accuracy across 40 applications with cross-device and cross-network transferability, on real operational cellular networks using open-source software. This is reproducible with moderate effort using the same class of tooling as Cheng \textit{et al.}~\cite{cheng2023watching}. We note that although these two works operate on bidirectional flow-level metadata (Table \ref{tab:pna-feasibility}, input type FLW), they differ from the flow-based works in Class 1 in one essential respect: rather than passively observing an already end-to-end-ciphered channel, both obtain flow-level metadata through relay-based interception. We therefore classify such flow-type inputs under the same criteria as control-plane/physical-layer inputs (Section \ref{subsec:class3}), rather than defaulting to Class 1 as we do for passively captured, still-encrypted flows.
    \item Wan \textit{et al.}~\cite{wan2024nr} and Wani \textit{et al.}~\cite{wani2024security} extend control-plane analysis to live 5G and NSA deployments, respectively. Their tools decode RRC and NAS signaling and associate UE activity with resource allocations, achievable on commercial 5G testbeds with calibrated SDR receivers.
    \item Jawne \textit{et al.}~\cite{jawne2025ai} and Zhang \textit{et al.}~\cite{zhang2025passive} exploit physical-layer information in 5G signals, such as RF fingerprints and uplink ACK/NACK behavior, to infer device presence or activity. Both have been experimentally validated on 5G NR, requiring signal synchronization and noise calibration but otherwise reproducible with standard research-grade SDR platforms.
    \item Ludant \textit{et al.}~\cite{ludant2024unprotected} extend control-plane analysis to the MAC and physical layers, showing that unprotected beam-reporting and other low-layer signaling can passively localize and track users with sub-cell precision. Demonstrated across commercial operators using standard downlink sniffing equipment, this attack requires no brute-force decoding and is reproducible with moderate effort.
    \item Li \textit{et al.}~\cite{li2026robust} present Crosshair. In its evaluated configuration, it avoids both physical-layer barriers identified in Section \ref{subsec:factors}: it observes the UE's uplink rather than the gNodeB's beamformed downlink, and it recovers channel occupancy through raw IQ energy detection rather than by decoding the RNTI-scrambled DCI. Its reception nevertheless depends on the UE's transmit pattern, power control, and distance, and NR uplink can itself be beamformed, particularly in FR2 \cite{3gpp38213, 3gpp38214}. With this caveat, it is the closest any surveyed work comes to Class 4 feasibility. We classify it as Class 3 because it still relies on SDR capture hardware and a custom research-grade cross-modal DL pipeline, rather than commodity hardware and turnkey open-source software (see Table \ref{tab:tools}). Should future work reproduce this side-channel across deployment configurations with commodity hardware and released tooling, it would be \textbf{the first attack} in our survey to qualify for Class 4.
\end{itemize}

\subsubsection{Class 4: Easily Reproducible} This class would include attacks that can be reproduced on 5G with minimal additional effort, using commodity hardware, readily available open source software, and without strong constraints on attacker positioning or signal directionality. In Wi-Fi and even LTE, examples of such attacks existed: basic website fingerprinting, simple traffic classification, and user/app identification could be performed with laptop-grade NICs and publicly available tools. 

However, across all the works surveyed, no attack meets these criteria in 5G. Even the most convincing 5G demonstrations, such as IMSI catching in NSA networks, PDCCH/DCI decoding for user presence, or application identification in controlled testbeds, required SDR hardware, custom protocol decoders, or non-commodity data collection pipelines. Signal directionality further complicates capture, since an attacker must obtain sufficient SINR for the selected transmission format, which depends on its position, antennas, and the channel. Likewise, software maturity remains low: although frameworks such as srsRAN exist, they require significant modification and lack turnkey support for 5G passive sniffing. 
In its evaluated configuration, the recent Crosshair attack \cite{li2026robust} avoided the downlink beam-selectivity and RNTI-decoding barriers, and it represents the most significant step toward Class 4 feasibility observed to date. Its reception remains configuration-dependent, however, and its hardware and software requirements still fall short of the commodity threshold this class demands.

\subsection{Discussion} \label{subsec:discussion}
The classification of prior works reveals clear trends in the feasibility of PNAs on 5G. Attacks that relied heavily on packet-level visibility or higher-layer traffic flows (e.g., TLS/HTTPS metadata, transport-level burst features) largely fall into \emph{Class 1} and \emph{Class 2}, since encryption and beamforming prevent external adversaries from reconstructing such data in 5G. In contrast, the attacks that remain feasible are those exploiting side-channels at the control plane or radio access layer. This includes works that decode PDCCH/DCI scheduling, measure timing signals, or analyze packet bursts indirectly observable at the air interface. These attacks form the bulk of \emph{Class 3}, showing that while they require non-trivial expertise and specialized SDR setups, they remain reproducible in 5G. Interestingly, no study reached \emph{Class 4}: even attacks demonstrated directly on 5G involve custom hardware, tuned decoders, or controlled conditions. The absence of \emph{``easily reproducible”} PNAs underscores that 5G has significantly raised the entry barrier compared to LTE and Wi-Fi, shifting the attack surface towards complex side-channels rather than straightforward traffic analysis.

A second important observation emerging from this study is the clear correspondence between the \emph{type of input data} used by prior attacks and their reproducibility on 5G. 
As summarized in Table~\ref{tab:pna-feasibility}, most \emph{Class~1} works rely on \textit{flows} as input, i.e., transport- or application-layer aggregates that are fully hidden behind PDCP encryption and GTP encapsulation in 5G. These dependencies on higher-layer visibility explain why such attacks are now infeasible. 
In contrast, the majority of \emph{Class~2} attacks operate on \textit{packet sequences}, using timing and size patterns as side channels. These features still exist conceptually in 5G; but recovering them requires extensive SDR-based demodulation and PDCP event reconstruction, hence their placement in the \emph{“reproducible with extensive efforts”} class. 
Finally, nearly all \emph{Class~3} works exploit \textit{control-plane messages} or \textit{physical-layer features} such as DCI scheduling, uplink feedback, or RF characteristics. These elements remain observable at the air interface and therefore represent the most practical input sources for modern PNAs. 

\noindent This strong alignment between input type and feasibility highlights a fundamental shift in adversarial observability: as encryption removes access to packets and flows, viable side channels have progressively \textbf{migrated downward} to the control and physical layers.

%Looking forward, this classification has important implications for B5G/6G. Since most \emph{Class 1} and \emph{Class 2} attacks are already infeasible in 5G, they are unlikely to reappear in more secure architectures~\cite{abdel2022security, alqwider2024combat}. The attacks that deserve closer attention are those in \emph{Class 3}, these represent the most reproducible threats today and may still be relevant if future systems do not fully obfuscate scheduling or timing information. However, with trends in B5G/6G towards randomized control channels, beam obfuscation, and AI-driven scheduling, it is reasonable to expect that even \emph{Class 3} attacks will become increasingly difficult, migrating toward \emph{Class 2} or even \emph{Class 1}. In this sense, a new classification table for B5G/6G as Table \ref{tab:pna-feasibility} is not strictly necessary: the current analysis already suggests that the feasibility of PNAs will continue to shrink, with adversaries needing ever more effort, expertise, and privileged access to reproduce them.
Looking forward, this classification has implications for B5G/6G, but they must be stated conditionally. Class 1 attacks depend on flow-level visibility that PDCP ciphering already removes in 5G, and there is no indication that B5G/6G will restore it \cite{abdel2022security, alqwider2024combat}. The attacks that deserve closer attention are those in Class 3, which exploit control-plane and physical-layer observables that remain exposed at the air interface. Whether they persist in B5G/6G depends on design choices that are not yet settled, such as whether control-channel scheduling and timing information remain observable.

Radio characteristics alone do not settle the question. B5G/6G will span a wide range of frequencies: from upper mid-band spectrum such as 7.125–8.4 GHz \cite{sampath2024imt2030}, through FR2 bands above 28 GHz already supported in NR \cite{3gpp381012}, to sub-THz bands. Higher frequencies and narrower beams increase path loss and spatial selectivity, but they do not by themselves make reception infeasible; reflection-assisted eavesdropping has been demonstrated against highly directional 60 GHz links \cite{steinmetzer2015periscopes}. Non-terrestrial links, whose footprints can cover large areas, and reconfigurable intelligent surfaces, which can redirect energy toward or away from a location, add further geometry-dependent cases.

For these reasons, we do not extend Table \ref{tab:pna-feasibility} to B5G/6G. Current evidence is insufficient to assign classes, and feasibility will depend on frequency, antenna configuration, geometry, and channel conditions rather than on the generation label alone.

%The hardware requirements summarized in Table \ref{tab:tools} scale with the capture task rather than with the network generation alone: control-channel observation remains within reach of mid-range SDRs, whereas full-carrier capture in FR1, and especially in FR2, requires substantially more capable equipment.
%The resource-barrier analysis in Section \ref{subsec:hardware-cost} (Table \ref{tab:tools}) can also be expressed as a concrete cost multiplier, which quantifies the entry-barrier claim made throughout this section: entry-level LTE passive-sniffing hardware costs under \$200, whereas comparable 5G NR capability requires \$500–\$2,000, a 2.5× to 10× increase, and B5G/6G prototype-grade hardware exceeds \$10,000, more than 50× the cost of an LTE-capable setup and 5× to 20× the cost of a 5G-capable one. This scaling factor holds independently of any specific attack technique, reinforcing that the shift we observe from Class 1/2 toward Class 3 feasibility, and the complete absence of Class 4, reflects a structural increase in the resource requirements for passive 5G/B5G eavesdropping, not merely the difficulty of any individual attack surveyed.

\begin{table*}[t]
\begin{center}
\caption{\small Comparison of the feasibility factors of PNAs in 5G and B5G/6G.}
%evaluation scale
%monitoring time
\label{tab:feas}
\footnotesize
\begin{tabular}{p{2.5cm} p{7.5cm} p{7cm}}

\hline 
\textbf{Aspect} & \textbf{5G} & \textbf{B5G / 6G} \\ 

\hline 

Signal Directionality & Beamformed data and control channels; decoding requires sufficient SINR via main lobe, side lobe, or multipath & Depends on frequency and array size; reception via side lobes, reflections, or RIS remains geometry-dependent \cite{steinmetzer2015periscopes} \\ 

Carrier Frequency & FR1 (e.g., 3.5 GHz) and FR2 (24.25–71 GHz) \cite{3gpp381012} & Upper mid-band (e.g., 7.125–8.4 GHz) \cite{sampath2024imt2030} to mmWave/sub-THz; path loss rises with frequency\\ 

SDR Requirements & Task-dependent: mid-range for SSB/CORESET 0, high-end for full-carrier capture & Task-dependent; wider bandwidths and arrays raise capture requirements \\ 

Tool Availability & Limited passive sniffers (custom implementations only) & No public sniffer stacks \\ 

Encryption / Integrity & IMSI/NAS protection, UP integrity optional & Stronger, integrated integrity/encryption \\ 

Packet/Flow Visibility & Link-layer only (no IP or payload access) & Same or more restricted (header and timing inference only) \\ 

Control-plane Leakage & DCI, RRC, MAC CE-based inferences possible \cite{wan2024nr, ludant20235g, ludant2024unprotected}& Unknown; depends on control-channel design \\ 

ML Attack Feasibility & Limited by noise and lack of ground truth & Not yet evaluated \\ 

Device Fingerprinting & Viable using RF features~\cite{jawne2025ai} &
Not yet evaluated\\ 

\textbf{Reproducibility} & \textbf{Challenging, but possible under specific conditions} & \textbf{Not yet demonstrated; no evidence of general infeasibility} \\ 

\hline

\end{tabular}
\end{center}
\end{table*} 

We summarize these findings in Table \ref{tab:feas} which outlines the technical and practical constraints that further limit the feasibility and reproducibility of PNAs. \textbf{First,} many 5G PNA studies assume clean, high-quality traces, but in real deployments, packet captures are often incomplete or degraded, especially under beam misalignment or high mobility, which dramatically reduces the reliability of ML-based classifiers. \textbf{Second,} beamformed transmission makes the SINR received by an eavesdropper sensitive to its position, antenna gain, and the surrounding channel, so passive capture success varies strongly with geometry and environment. \textbf{Third,} control-plane leakage via DCI decoding or RRC signaling may provide useful side-channel information, but it's increasingly recognized as a substitute for DPI, carrying potential risks related to traffic censorship or user fingerprinting even under encryption \cite{rupprecht2019breaking}. \textbf{Fourth,} edge deployments and V2X scenarios, characterized by rapid node movement and dynamic channels, introduce further challenges; propagation conditions vary dramatically, making attacker alignment and signal acquisition at the edge an open research question. \textbf{Fifth,} most existing studies assume idealized lab settings, limiting their generalizability; this echoes the device- and vendor-generalization limitations we noted earlier in Section \ref{subsec:challenges5g}, where models trained on one device or chipset were shown to generalize poorly to others~\cite{aceto2019mobile, aceto2020toward, ismailaj2021deep}. \textbf{Finally,} at the B5G/6G frontier, no open passive sniffer stacks exist and most experimentation is confined to proprietary testbeds, so no PNA has yet been demonstrated against B5G/6G systems. This absence reflects the current maturity of tooling rather than a demonstrated physical barrier: higher-frequency bands and tighter beam steering raise path loss and spatial selectivity, but reflection-assisted interception of directional links has been shown to be possible \cite{steinmetzer2015periscopes}.

\subsection{Emerging Attack Surfaces Beyond the Current Scope} \label{subsec:aiot}
The feasibility analysis above is grounded in the 44 works meeting our selection criteria (Section \ref{subsec:data-selection}), which target conventional 5G smartphone-class traffic. 3GPP Release-18 and Release-19 introduce several new device classes and mechanisms that fall outside this scope but raise related passive-attack questions worth addressing explicitly, both to bound our contribution honestly and to identify concrete future work.

\subsubsection{Ambient IoT and energy-harvesting devices}
3GPP Release-19 is standardizing Ambient IoT (AIoT), battery-free or energy-harvested devices with severely constrained compute and radio budgets, with traffic models and use cases defined in TR 22.840~\cite{3gpp22840} and RAN feasibility studies in TR 38.848~\cite{3gpp38848}; corresponding lightweight security requirements are under development in TS 33.369~\cite{3gpp33369}. For a passive attacker, AIoT introduces a double-edged characteristic. On one hand, the intermittent, low-duty-cycle, and often unencrypted-by-design signaling proposed for AIoT (to conserve energy) may create a more observable side-channel than conventional NR traffic: reduced cryptographic overhead directly implies reduced confidentiality, and TS 33.369's stated challenge is precisely how to add protection without breaking the energy budget. On the other hand, AIoT's minimal, event-driven traffic (short bursts, no persistent session) may not carry the rich timing/size features that the ML-based PNAs surveyed in Section \ref{sec:methodology} rely on, most of the 44 works assume sustained flows or repeated control-channel activity, neither of which is guaranteed for a device that transmits only sporadically. We consider AIoT passive-attack feasibility an open question that cannot be answered from currently published work, since, to our knowledge, no PNA has yet been demonstrated or evaluated against an AIoT-class device; we flag this explicitly as future work rather than speculate further.

\subsubsection{Network power-saving mechanisms}
Release-18 power-saving features, including extended DRX (eDRX) and reduced monitoring for RedCap devices, change when a device is reachable and how often its signaling security context is refreshed. This has a direct, testable implication for the control-plane-based PNAs in Class 3 of our taxonomy (Table \ref{tab:pna-feasibility}): attacks such as Ludant et al.~\cite{ludant20235g} and Budykho et al.~\cite{budykho2023fine} rely on continuously observing scheduling or RRC signaling to build a trace; eDRX's longer sleep cycles would reduce the observable signaling window per device, likely lowering attack success rates and moving such attacks toward Class 2. Conversely, longer, more predictable wake cycles could make timing-based correlation (i.e., linking a device's wake pattern across sessions) an easier side-channel than it is for continuously connected devices today. Evaluating this trade-off empirically is a natural extension of the real-5G-testbed validation we outline in our future work (Section \ref{sec:conclusion}), but is outside the current literature-based scope.

\subsubsection{Physical-layer security beyond PDCP/NAS encryption}
3GPP intentionally leaves several lower-layer control messages, e.g., MAC Control Elements and RLC/PHY procedure signaling, unencrypted and unauthenticated to meet strict timing constraints~\cite{3gpp38300, 3gpp33501}. Ludant et al. \cite{ludant2024unprotected} have since demonstrated this experimentally, showing that unprotected MAC CE and beam-reporting signaling suffices to passively localize and track users (Section \ref{subsec:position-tracking}). This is directly relevant to the control-plane/physical-layer input class already central to our Class 3 findings (Section \ref{subsec:class3}): the same design choice that we identify as making DCI/RRC-based PNAs reproducible (unencrypted scheduling metadata) is, from a physical-layer-security (PLS) perspective, a known and studied vulnerability class in its own right, extending beyond identification/tracking (our scope) to include control-signaling manipulation, spoofing, and throughput-reduction attacks. We view PLS-based confidentiality techniques, e.g., artificial noise, coding-based secrecy, waveform design, as a promising countermeasure direction for the very attacks this survey catalogs, and note this as future work rather than incorporating a full PLS threat review, which is already the subject of dedicated surveys we cite in Table \ref{tab:survey_comparison} (Devi et al.~\cite{DEVI2025100891}, Harvanek et al.~\cite{s24175523}).

\subsubsection{Positioning and sensing security}
3GPP Release-16 and Release-17 standardize network-based positioning methods (multi-round-trip-time and angle-based techniques), with requirements in TS 22.261~\cite{3gpp22261}. These are architecturally distinct from the geolocation attacks surveyed in Section \ref{subsec:position-tracking} (which infer position passively from side-channel leakage of otherwise unrelated traffic, e.g.,~\cite{yang2016passive, ateniese2015no, kotuliak2022ltrack}) in one important respect: standardized positioning is an active, network-cooperative service, so attacks against it (spoofing or manipulating RTT/angle measurements to produce incorrect location output) are properly classified as active integrity attacks on a location service, not PNAs in the sense we define in Section \ref{subsec:attack-types}. That said, an external passive observer with the RF and protocol capability discussed in Section \ref{subsec:factors} could, in principle, silently monitor exchanged positioning reference signals in the same way it monitors PDCCH scheduling information, converting what is designed as a cooperative measurement channel into a fourth passive side-channel; we are not aware of any published work evaluating this specific scenario, and we flag it as a concrete gap for future study rather than one this survey's 44-work sample can currently support.

\section{Conclusion and future work} \label{sec:conclusion}
This survey systematically analyzed the current landscape of passive network attacks in the context of 5G and beyond-5G cellular networks. We reviewed 44 peer-reviewed works, covering attack goals such as traffic classification, app and video identification, and user localization. We classified these works based on their input types (packet sequences, flows, or physical/control-plane data) and analyzed their applicability to next-generation networks.

Our main research question was to determine how feasible and reproducible existing PNAs are when applied to 5G and B5G deployments, and therefore ease the public concerns about the security and privacy implications of 5G architectures. The comparative analysis shows that most classical attacks, such as flow-based fingerprinting, are no longer reproducible by an external passive observer in 5G, since PDCP ciphering and multiplexing remove the required visibility. Packet-sequence attacks remain possible only with extensive SDR-based reconstruction. Additional obstacles include the need for task-appropriate SDR hardware and the limited maturity of open-source 5G sniffing tools. Nevertheless, recent studies demonstrate that a subset of PNAs, exploiting control-plane leaks (e.g., through DCI, RRC, or MAC CE analysis) and physical-layer side channels (e.g., RF fingerprinting or uplink occupancy), remains feasible in 5G, though only with non-trivial expertise and equipment. Relay-based attacks \cite{cheng2023watching, wang2020automatic} further show that an adversary able to interpose itself between the UE and the network recovers flow-level metadata with high accuracy, although such attacks fall outside the strictly passive threat model.

In other words, 5G does not eliminate PNAs, but it confines them to research-grade or well-equipped adversaries. The closest approach to commodity feasibility so far, Crosshair \cite{li2026robust}, exploits the uplink in a configuration-dependent way. For B5G/6G, no passive attack has yet been demonstrated and no public sniffing tools exist, but available evidence does not establish general practical infeasibility. Feasibility will depend on frequency, antenna configuration, geometry, and channel conditions, and directional links have already been shown to be vulnerable to reflection-assisted eavesdropping \cite{steinmetzer2015periscopes}.

\paragraph*{Limitations}
This work is based on a literature-driven feasibility study. We have not yet performed empirical validation in commercial or operational 5G environments due to limited availability of antennas and suitable passive hardware. Additionally, many published attacks rely on clean or synthetic traces, whereas real-world sniffing introduces noise, packet loss, and multi-flow mixing; factors that are often unaddressed in prior evaluations.

\paragraph*{Future works}
Our next steps involve testing selected passive attacks on a real 5G testbed. This includes evaluating:
\begin{itemize}
    \item The impact of beam misalignment and mobility on trace capture,
    \item The minimum proximity required for successful signal interception, and
    \item The reproducibility of known classification models in degraded conditions.
\end{itemize}

In parallel, we plan to explore probabilistic modeling tools such as Hidden Markov Models and finite state machines for application classification under link-layer encryption, where traditional flow-based ML fails due to lack of packet headers. These models may offer resilience to noise and enable inference from sparse metadata.

In the longer term, our work may contribute to improving both attack resilience in 5G/B5G systems and autonomous traffic classification in data centers and edge networks, including developing DPI alternatives that respect encryption boundaries while ensuring performance and QoS.

Ultimately, this survey lays the groundwork for more realistic, implementation-based evaluations of passive inference risks in next-generation wireless networks. If attacks fail under real-world constraints, 5G can be shown to offer a measurable privacy advantage over LTE. If not, our findings may inform future standards and mitigation strategies.
%\appendices

\section*{Acknowledgment}
Claude Opus 5.5 has been used to recreate the figures of this paper 

%\section{bib}
\bibliographystyle{IEEEtran}
\bibliography{references}
\newpage
\begin{IEEEbiography}[{\includegraphics[width=1in,height=1.25in,clip,keepaspectratio]{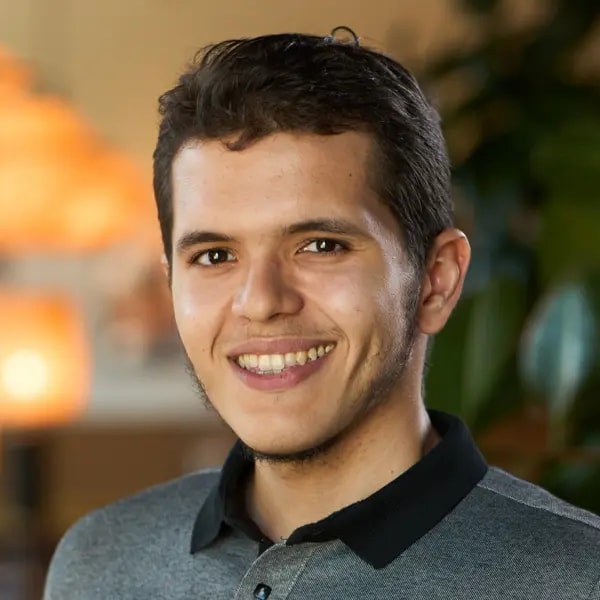}}]{\textbf{Atmane A. Mansour Bahar}} (Graduate Student Member, IEEE) received his degree in
Computer-Systems Engineering from the École Nationale Supérieure
d’Informatique (ESI ex. INI), Algiers, Algeria
For his graduation project, he explored leveraging Federated Learning with Graph Neural Network (GNN)-based model for real-time detection of Advanced Persistent
Threats (APTs). He then started pursuing his PhD in Encrypted Traffic Classification at Chalmers University of Technology since Mid 2025, where he is exploring solutions to Side-Channel attacks related to Video Fingerprinting.
His research interests include AI-based intrusion detection systems, privacy-preserving technologies, and the security of computer systems and networks.

\end{IEEEbiography}
\begin{IEEEbiography}[{\includegraphics[width=1in,height=1.25in,clip,keepaspectratio]{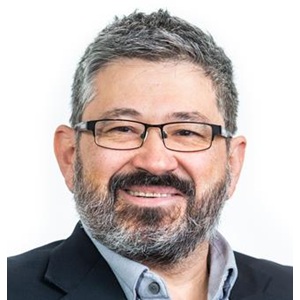}}]{\textbf{Andrés Alayón Glazunov}} (Senior Member, IEEE) was born in Havana, Cuba. He received the M.Sc. (Engineer-Researcher) degree in physical engineering from Peter the Great St. Petersburg Polytechnic University (Polytech), Saint Petersburg, Russia, in 1994, the Ph.D. degree in electrical engineering from Lund University, Lund, Sweden, in 2009, and the Docent (Habilitation) degree in antenna systems from the Chalmers University of
Technology, Gothenburg, Sweden, in 2017.
From 1996 to 2005, he held various research and specialist positions with the telecom industry, e.g., Ericsson Research, Telia Research, and TeliaSonera; Stockholm, Sweden. From 2001 to 2005, he was the Swedish Delegate with the European Cost Action 273, and from 2018 to 2020, he was the Dutch Delegate with the European Cost Action IRACON. He has been one of the pioneers in producing the first standardized OTA measurement techniques for 3GPP, and devising novel OTA techniques, e.g., the random-LOS and the hybrid antenna characterization setups. He has contributed to, or initiated various European research projects, e.g., more recently, is3DMIMO, WAVECOMBE, 5VC, and Build-Wise projects under the auspices of the H2020 European Research and Innovation Program. He also contributed to the international 3GPP and the ITU standardization bodies. From 2009 to 2010, he held a Marie Curie Senior Research Fellowship with the Centre for Wireless Network Design, University of Bedfordshire, Luton, U.K. From 2010 to 2014, he held a postdoctoral position with the Electromagnetic Engineering Laboratory, KTH-Royal Institute of Technology, Stockholm. From 2014 to 2018, he was an Assistant Professor with the Chalmers University of Technology. From 2018 to 2023, he was an Associate Professor with the Department of Electrical Engineering, University of Twente, Enschede, The Netherlands. He is currently a Senior Associate Professor with the Department of Science and Technology, Linköping University, Norrköping, Sweden. He is the author of more than 150 scientific and technical publications. He is the coauthor and the coeditor of the text book LTE-Advanced and Next Generation Wireless Networks-Channel Modelling and Propagation (Wiley, 2012). His current research interests include, but are not limited to, mmWave array antenna design, MIMO antenna systems, electromagnetic theory, fundamental limitations on antenna-channel interactions, radio propagation channel measurements, modeling and simulations, wireless performance in the built environment, and the OTA characterization of antenna systems and wireless devices for communication and sensing.
\end{IEEEbiography}

\begin{IEEEbiography}[{\includegraphics[width=1in,height=1.25in,clip,keepaspectratio]{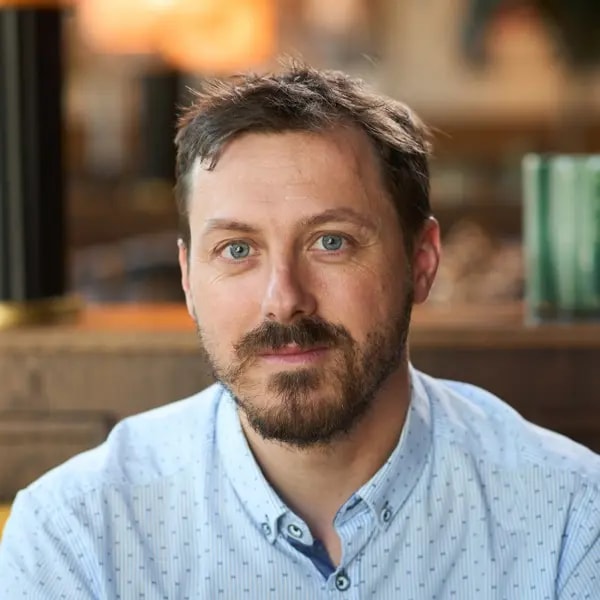}}]{\textbf{Romaric Duvignau}} (Senior Member, IEEE) received the Ph.D. degree in computer science from the University of Bordeaux (LaBRI), France, in 2015.
He is currently an Associate Professor and Head of the Network and Systems Unit at the Computer Network and Systems Division, Chalmers University of Technology. 
He was previously affiliated with Aix-Marseille University (LIF). His research interests include data stream processing, edge computing, p2p energy trading, continuous distributed monitoring, and discovery algorithms for assignment problems.
\end{IEEEbiography}

%\EOD

\end{document}